\documentclass[fleqn,usenatbib]{mnras}

\usepackage{newtxtext,newtxmath}
\usepackage{ulem}

\usepackage[T1]{fontenc}

\DeclareRobustCommand{\VAN}[3]{#2}
\let\VANthebibliography\thebibliography
\def\thebibliography{\DeclareRobustCommand{\VAN}[3]{##3}\VANthebibliography}

\usepackage{graphicx}	

\title[Gas Phase Metals in FIRE-2 Dwarfs]{{Spatially Resolved Gas-phase Metallicity in FIRE-2 Dwarfs: Late-Time Evolution of Metallicity Relations in Simulations with Feedback and Mergers}} 

\author[L. E. Porter et al.]{
Lori E. Porter,$^{1}$\thanks{E-mail: leport01@louisville.edu}
Matthew E. Orr,$^{2,3}$
Blakesley Burkhart,$^{2,3}$
Andrew Wetzel,$^{4}$
Xiangcheng Ma,$^{5}$ \newauthor
Philip F. Hopkins,$^{6}$ 
{Andrew Emerick},$^{6,7}$
\\
$^{1}$Department of Physics and Astronomy, University of Louisville, 102 Natural Sciences Building, Louisville, KY 40292, USA\\
$^{2}$Department of Physics and Astronomy, Rutgers, The State University of New Jersey, 136 Frelinghuysen Rd, Piscataway, NJ 08854, USA\\
$^{3}$Center for Computational Astrophysics, Flatiron Institute, 162 Fifth Avenue, New York, NY 10010, USA\\
$^{4}$ Department of Physics and Astronomy, University of California, Davis, CA 95616, USA \\
$^{5}$ Department of Astronomy and Theoretical Astrophysics Center, University of California Berkeley, Berkeley, CA 94720 \\
$^{6}$TAPIR, California Institute of Technology, 1200 E. California Blvd., MC 350-17, Pasadena, CA 91125, USA\\
$^{7}$Carnegie Observatories, 813 Santa Barbara Street, Pasadena, CA 91101, USA\\
}

\date{Accepted XXX. Received YYY; in original form ZZZ}

\pubyear{2022}

\begin{document}
\label{firstpage}
\pagerange{\pageref{firstpage}--\pageref{lastpage}}
\maketitle

\begin{abstract}
We present an analysis of spatially resolved gas-phase metallicity relations in five dwarf galaxies ($M_{halo} \approx 10^{11} M_\odot$, $M_\star \approx 10^{8.8}-10^{9.6} M_\odot$) from the FIRE-2 (Feedback in Realistic Environments) cosmological zoom-in simulation suite, which include an explicit model for sub-grid turbulent mixing of metals in gas, near $z\approx 0$, over a period of 1.4 Gyrs, and compare our findings with observations. While these dwarf galaxies represent a diverse sample, we find that all simulated galaxies match the observed mass-metallicity (MZR) and mass-metallicity gradient (MZGR) relations. We note that in all five galaxies, the metallicities are effectively identical between phases of the interstellar medium (ISM), with 95$\%$ being within $\pm$0.1 dex between various ISM phases, including the cold and dense gas ($T < 500$~K and $n_{\rm H} > 1$ cm$^{-3}$), ionized gas (near the H$\alpha$ $T \approx 10^4$ K ridge-line), and nebular regions (ionized gas where the 10 Myr-averaged star formation rate is non-zero). We find that most of the scatter in relative metallicity between cold and dense gas and ionized gas/nebular regions can be attributed to either local starburst events or metal-poor inflows. We also note the presence of a major merger in one of our galaxies, m11e, with a substantial impact on the metallicity distribution in the spatially resolved map, showing two strong metallicity peaks and triggering a starburst in the main galaxy.
\end{abstract}

\begin{keywords}
galaxies: dwarf -- galaxies: ISM -- ISM: kinematics and dynamics -- galaxies: evolution -- ISM: abundances
\end{keywords}



\section{Introduction}

Dwarf galaxies are one of the most important components of galactic evolution. They are the lowest-mass galaxies, but are the most abundant type and form the bottom of the hierarchy of galactic evolution. At low masses, the effects of individual starburst events are particularly pronounced since individual events can disturb the entire existing gas reservoir in the galaxy. Due to their abundance and susceptibility to feedback, dwarf galaxies are ideal testbeds for analyzing various forms of enrichment and feedback physics \citep[][]{Tolstoy2009, Simon2019}.

Gas-phase metallicity enrichment is essential to our understanding of galactic formation and evolution. The metal enrichment that follows supernovae (SNe) surrenders information about star formation history \citep{Tolstoy2009, Sacchi2016} and the assembly history of galaxies \citep{Sawala2010, Pawlik2013, Hirschmann2016}. When metals are distributed throughout the interstellar medium (ISM) by SNe marking the death of high mass stars, these metals constrain star formation and trace feedback from massive stars and SNe. These stellar yields are also components that influence galactic evolution.  
In particular, the gas-phase oxygen abundance in ionized gas is significant, as oxygen is the most abundant metal element and a primary coolant of the ISM \citep{Draine2011}. It produces strong optical emission lines when ionized and is used as a tracer of  metallicity in the ISM \citep[e.g.,][]{Sanchez2019}. 

One of the most well-known relationships for galaxy evolution and oxygen abundance is the mass-metallicity relation (MZR). This relation is the strong correlation between stellar mass and both the gas-phase metallicity and stellar metallicity \citep{Gallazzi2005, Lee2008, Kirby2013, Jimmy2015, Ma2016, Hidalgo2017}; more massive galaxies tend to have higher metal enrichment than their low-mass counterparts (i.e, dwarf galaxies). \citet{Kirby2013} rationalizes the MZR by arguing that the deeper potential wells of more massive galaxies more easily retain metals, as they produce them through in-situ star formation. Dwarf galaxies, such as the ones in this study, lack the mass, and therefore the gravity, to resist feedback mechanisms from expelling their metals. The increased effectiveness of such feedback allows for the gas-phase metal reservoirs to be diluted in samples of dwarf galaxies as compared to their more massive counterparts. In addition, lower-mass galaxies are typically more gas-rich, resulting in diluted metals in the case that they are retained, such that even a closed-box model would show the MZR. 

Star formation efficiency also has a role to play in the MZR, as galaxies that have increased star formation rates (SFRs) will reach a higher stellar mass and metallicity enrichment. The stellar mass-halo mass relation presented in \citet{Behroozi2013} and \citet{Moster2013} suggests that lower SFR efficiency in dwarf galaxies results in less metal enrichment. \citet{Hopkins2013d} also finds that the presence of mergers can lower star formation after producing starbursts. For the masses relevant to this study ($ M_* \approx  10^{8.75}-10^{9.75} M_\odot$; see Figure~\ref{fig:SFR_Mstar_overtime}), the MZR can typically be characterized by a power law. \citet{Lee2006} supports this, confirming that a tight correlation exists among local dwarf galaxies with stellar masses from $10^6-10^{11} M_{\odot}$. 

In addition to the MZR, gas-phase metallicity gradients in dwarf galaxies can contribute insight into galaxy assemblies and provide information as to how ordered the systems are. This is because metals are not uniformly distributed throughout a galaxy \citep{Shaver1983}. Instead, feedback has a predominant role to play in the distribution of metals throughout a galaxy \citep{Pilkington2012, Hemler2021, Patel2021}. For example, \citet{Gibson2013} uses cosmological hydrodynamical simulations to investigate how energy feedback affects the distribution of metals in disk galaxies, finding that flat and negative gradients are common in galaxies with strong feedback redistributing metals. In contrast, simulations with weaker feedback tend to produce steeper gradients. \citet{Hemler2021} also finds that many TNG50 galaxies ($10^{10}M_{\odot} \leq M_{*} \leq 10^{10.5}$) also exhibit negative gradients around $z\approx 0$, although they note that TNG50 gradients tend to be steeper at higher redshifts than similar galaxies in the FIRE (Feedback In Realistic Environments) simulations.

\citet{Searle1971} introduces the concept of using spatial metallicity gradients as a function of radius \citep{Ho2017, Kreckel2019, Bellardini2021}. \citet{Sharda2021} states the metallicity gradient measured in dex per kpc has either one of two properties at stellar mass of $M_{*} \sim 10^{10-10.5} M_{\odot}$: it will be independent of stellar mass up to this point, then flatten toward zero at higher stellar masses \citep{Hemler2021}, exactly the opposite as found in FIRE-2 \citep{Mercado2021}, or it will have a mild curvature with flat gradients on either side \citep[e.g.,][]{Belfiore2017}. Further investigation into the physical cause of the difference in gradients, similar to the MZR,  can provide clues as to how feedback and turbulence can affect the population of galaxies at a lower mass, and how dwarf galaxies lose their metals \citep{DekelSilk1986, Dalcanton2007, Burkhart2010, Ma2016}. In addition, the expectation that MW-mass disk galaxies exhibit negative radial metallicity gradients provides evidence of inside-out galaxy formation \citep{Mo1998, Pilkington2012, Sharda2021}. This is found in \citet{Magrini2016}'s observations and \citet{Bellardini2021}'s azimuthal analysis in the FIRE-2 m12 galaxies (MW and M31-mass). However, as found in \citet{Mercado2021}'s comparison between FIRE-2 and Local Group dwarfs, dwarf galaxies behave differently due to their usually well-mixed ISM. Dwarf galaxies have the tendency to exhibit no radial gradients, but this varies across galaxies of similar masses \citep{Belfiore2017, Ma2017, Escala2018, Mercado2021}.

Cosmological zoom-in simulations are an excellent choice to study the metallicity distributions in dwarf galaxies. Several studies in this field have used one-zone models \citep{Lanfranchi2003, Lanfranchi2007, Lanfranchi2010, Yin2011}, which are faster than hydrodynamical simulations, but rely on more simplistic models of feedback. While these results can effective at explaining the phenomena occurring within the galaxies, it can be ineffective at capturing the true cosmological context surrounding a specific galaxy (e.g., the highly variable rates of mergers or inflows) and the non-linear interactions between stellar feedback within galaxies and their surroundings \citep{Ma2016, Escala2018}. This makes predicting metallicity gradients difficult in such models (they are necessarily "sub-grid" with a single zone), and they fail to address temporal scatter in the MZR and gradients. It is important to reproduce observed stellar masses, SFRs, and metallicities in dwarf galaxies, which requires the increased dynamic range found in such simulations \citep{Ma2016, Hemler2021}. 

This paper combines the powerful attributes of dwarf galaxies, the FIRE-2 (Feedback In Realistic Environments) simulations, and metal enrichment to form an analysis on gas-phase metallicity relations at this galaxy mass range. By combining all of these tools, we are effectively able to resolve the internal structures of galaxies while maintaining realistic cosmological context, using FIRE's ability to recreate a resolved multi-phase ISM, star formation, stellar feedback, and other forms of physics that result in successfully reproducing observations. In this paper we analyze five dwarf galaxies from the FIRE-2 simulations near z$\sim0$. We complete our analysis of their properties by creating spatially resolved maps in the cold and dense gas  and ionized gas (comparable to observed molecular gas and nebular emission tracers, respectively), progress to discussing the MZR, radial profiles and gradients, and MZGR, and analyze the variation in enrichment between the selected ISM phases. Finally, we make note of the presence of a merger in one of our simulated galaxies (denoted below as \textbf{m11e}), studying this simulation in slightly more detail. This paper is then concluded by a discussion and restatement of our findings compared to existing literature and observations.

\section{Simulations and Methods}

\begin{figure}
	\includegraphics[width=\columnwidth]{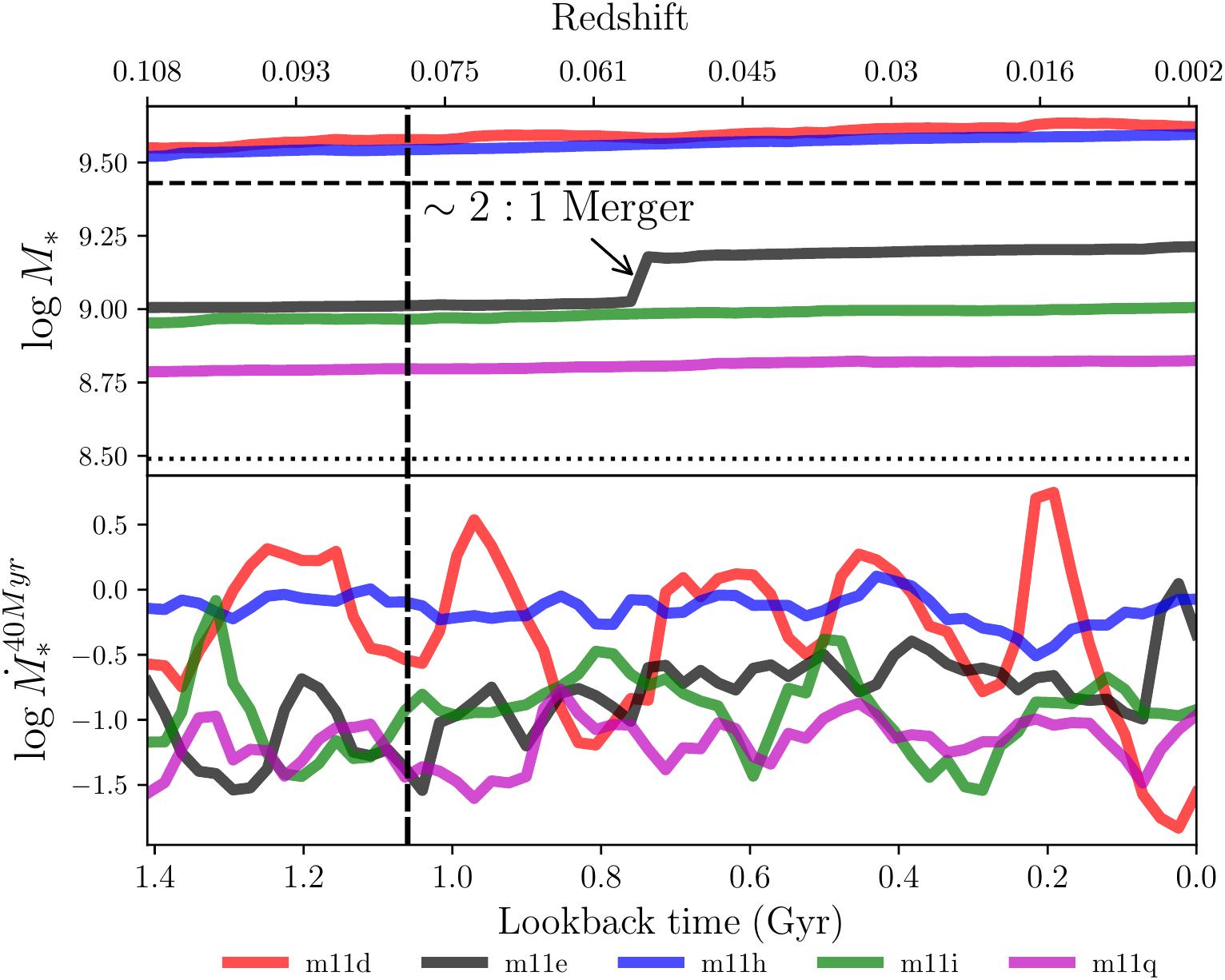}
    \caption{ \textit{Top panel}: Stellar masses in each snapshot for all five FIRE-2 dwarf galaxies. Stellar masses exhibit a slight steady growth over time. m11e (black) is an exception, with disturbances due to a merger. SMC/LMC masses from \citet{Bekki2009} and \citet{VanderMarel2002} for reference are denoted by the horizontal dotted and dashed black lines, respectively. 
    \textit{Bottom panel}: Star formation rates averaged over the last 40 Myrs for each snapshot. m11d is clearly bursty with variations over 2 dex. m11e has a merger-triggered starburst around $t_{lookback}=0$ Gyrs. m11i is somewhat bursty with variations over 0.5 dex. m11h and m11q are fairly smooth with recent star formation history. 
    Thickly dashed vertical line denotes $t_{lookback}=0.36$ Gyrs, which corresponds to the particular snapshot shown in Figures ~\ref{fig:allgals_imshows} and \ref{fig:m11e_m11h_imshows}.}
    \label{fig:SFR_Mstar_overtime}
\end{figure}

\begin{table}\caption{Summary of $z\approx0$ properties of the FIRE-2 dwarf galaxies used in this work}\label{table:galprops}
\begin{tabular}{lccccc}
\hline
Name & $\log(\frac{M_\star}{{\rm M_\odot}})$ & $\log(\frac{M_{\rm gas}}{{\rm M_\odot}})$ & $\frac{R_{\star,1/2}}{\rm kpc}$& $\frac{R_{\rm gas,1/2}}{\rm kpc}$ & $\frac{v_c}{\rm km/s}$\textsuperscript{\textdagger}\\ 
\hline
m11d$^\star$ & 9.6 & 9.5 & 7.0 & 11.4 & 85.6 \\
m11e & 9.2 & 9.3 & 3.9 & 6.6 & 82.7 \\
m11h & 9.6 & 9.6 & 4.1 & 6.3 & 97.7 \\
m11i & 9.0 & 9.2 & 3.8 & 3.4 & 53.1 \\
m11q & 8.8 & 9.2 & 2.6 & 4.2 & 68.8 \\
 \hline
\multicolumn{6}{l}{Note: all quantities measured within a 30~kpc cubic aperture.}\\
\multicolumn{6}{l}{\textsuperscript{\textdagger}Circular velocities evaluated at $R_{\rm gas,1/2}$.}\\
\multicolumn{6}{p{0.9\columnwidth}}{$^\star$m11d is significantly disrupted at $z\approx 0$ from a recent starburst (see, \emph{e.g.}, Fig.~\ref{fig:allgals_imshows}), as such its radius (and $v_c$ at that radius) is poorly defined.}\\
\multicolumn{6}{p{0.9\columnwidth}}{All galaxies here are introduced by \citet{El-Badry2018}, with the exception of m11q, which is introduced by \citet{Hopkins2018:fire}.} \\
\end{tabular}
\end{table}

We make use of five dwarf galaxies from the FIRE-2 cosmological zoom-in simulation suite \citep{Hopkins2018:fire} in our analysis here.  These galaxies have halo masses $\sim$$10^{11} M_\odot$, and stellar masses spanning from $\sim$$2\times$ the SMC mass to slightly more massive than the LMC (see Table~\ref{table:galprops} for a summary of the $z\approx 0$ galaxy masses, sizes and rotational velocities).  We use approximately 60 snapshots from each simulation, spaced $\sim25$~Myr in time (for a roughly 1.4 Gyr total period analyzed) near $z \approx 0$.  Figure~\ref{fig:SFR_Mstar_overtime} shows the evolution of the stellar mass ($M_\star$) and 40 Myr-averaged star formation rate ($\dot M_\star^{\rm 40 Myr}$) of the five galaxies over the epoch analyzed, with estimates for the stellar mass of the SMC and LMC included for comparison \citep{Bekki2009, VanderMarel2002}. 

We generate mock observational maps from these snapshots with the same method as \citet{Orr2020}: we project the dwarf galaxies face-on using the net angular momentum vector of the star particles within a 3-D stellar half-mass radius, and binning star particles and gas elements into square pixels with side-lengths (i.e., ``pixel sizes”) 250 pc. The maps are 30 kpc on a side and integrate gas and stars within $\pm$15 kpc of the identified galactic mid-plane.  This cube includes all of the galaxy body and much of the surrounding $\sim 10^4$ K gas halo for all five of the dwarf galaxies that we map.  One simulation however, \textbf{m11e}, undergoes a major ($\sim$2:1 $M_*$) merger in the analysis period; and to capture the interaction of the two warm gas halos surrounding the main galaxy and its companion, we expand the size of the mapping cube to 60 kpc on a side (and $\pm$30 kpc along the line-of-sight).

A detailed presentation of the star formation prescription, feedback physics, and enrichment processes used in these simulations can be found in \citet{Hopkins2018:fire}, however we briefly reiterate some of the relevant implementations here.  These runs all have minimum baryonic particle/element masses of $m_{b, {\rm min}} = 7100$ M$\odot$, adaptive force softening (with minimum softening lengths $<$1~pc), and a 10 K gas temperature floor.  As the softening lengths are adaptive, we point out that the median softening length for the gas elements in these dwarf galaxies is $h \approx 90$~pc for $n > 1$ cm$^{-3}$ (and $h \approx 50$~pc for $n > 10$ cm$^{-3}$). The spatially resolved maps that we produce in our analysis have a pixel size of 250 pc, which more than adequately resolves the cold and dense turbulent gas structures (these have softening lengths $\ll$50 pc).  The warm diffuse ionized gas in the galactic outskirts is marginally resolved at this pixel size, but our analysis focuses primarily on the better (spatially) resolved gas within the primary body of these dwarf galaxies, and this does not qualitatively affect our interpretations of the surrounding warm gas halos.

In these simulations, star formation proceeds in dense ($n >10^3$ cm$^{-3}$), molecular (following the scalings of \citealt{Krumholz2011}), self-gravitating (viral parameter $\alpha_{\rm vir} < 1$) and Jeans-unstable (below the smoothing scale) gas, at a rate of $\dot\rho_\star = \rho_{\rm gas}/t_{\rm ff}$ (where $t_{\rm ff}$ is the free-fall time of the gas element).  The resulting star particles are single stellar populations, with a single age, metallicity, and mass.  Feedback physics in the forms of supernovae, stellar mass loss (OB/AGB-star winds), photoionization and photoelectric heating, and radiation pressure are explicitly included.  There are no black holes (or their attendant feedback/accretion processes) in these simulations.

As we focus heavily on the metal reservoirs of these dwarf galaxies, we summarize here the abundance/enrichment implementation of FIRE-2 (again, see \citealt{Hopkins2018:fire} for more detail): nucleosynthetic yields from core-collapse SNe taken from \citet{Nomoto2006}, Type-Ia SNe yields from \citet{Iwamoto1999}, and stellar wind yields (from O, B, and AGB stars) from \citet{Wiersma2009}.  These simulations also include a sub-grid metal diffusion model, accounting for turbulent mixing in unresolved eddies \citep{Su2017, Escala2018}.  The diffusion term smooths the abundance distribution following the prescription of \citet{Shen2010}, resulting in a more realistic distribution of gas phase (and consequently stellar) metallicities \citep{Escala2018}. \citet{Escala2018} further elaborates and concludes that star-forming gas in FIRE-2 dwarf galaxies is typically well-mixed at any fixed time, and the enrichment over time leads to a stellar metallicity distribution that matches observations, which is only achieved when simulations model metal mixing.

We produce proxies for observational metallicity tracers by mapping the ionized gas near the H$\alpha$ $T \approx 10^4$ K ridge-line (specifically, we identify this gas in a band of $|\log T - 4.05 | < 1/6$), as well as cold and dense gas ($T < 500$~K and $n_{\rm H} > 1$ cm$^{-3}$).  We also generate an analogue of observational measures of recent SFRs by calculating the 10 Myr-averaged SFR in the pixels.  To do so, we find the mass of all age $<10$ Myr star particles in the pixel, and correct for stellar evolutionary effects using {\scriptsize STARBURST99} \citep{Leitherer1999}.  We choose this time interval because of its \emph{approximate} correspondence with the timescales traced by recombination lines like H$\alpha$ \citep{Kennicutt2012, Velazquez2020}, and to associate ionized gas near $T \approx 10^4$ K with star-forming regions as a proxy for identifying HII region nebulosity (FIRE-2 uses {\scriptsize STARBURST99} predictions for the ionizing photon budget of star particles).  We also calculate the 40 Myr-averaged SFR in the pixels, for the purposes of associating metallicity scatter with the approximate timescale of the duration of supernova feedback (and their enrichment). 

In comparing with observations, we identify the warm ionized gas column (WIM) and select gas elements with $|\log T - 4.05 | < 1/6$, regardless of density.  In our analysis, we associate this gas reservoir with diffuse warm ionized gas halos surrounding (and to an extent throughout) these galaxies and `HII regions' in star-forming clouds (i.e., pixels where the 10 Myr-averaged star formation rate is non-zero).  We compare our `HII region' columns and metallicities ($Z_{\rm HII}$) with nebular emission observations of, e.g., [OIII].  Lastly, we also identify the ``cold and dense'' gas ($Z_{\rm CD}$ throughout), having $T < 500$~K and $n_{\rm H} > 1$ cm$^{-3}$.  We gas relate this gas in the simulations with that traced by cold dust or CO.  We present the total, oxygen and iron gas-phase metallicities throughout our analysis, but focus mostly on the total metallicity of the `HII regions’ and cold and dense gas.

\section{Analysis and Results}

\begin{figure*}
	\includegraphics[width=0.9\textwidth]{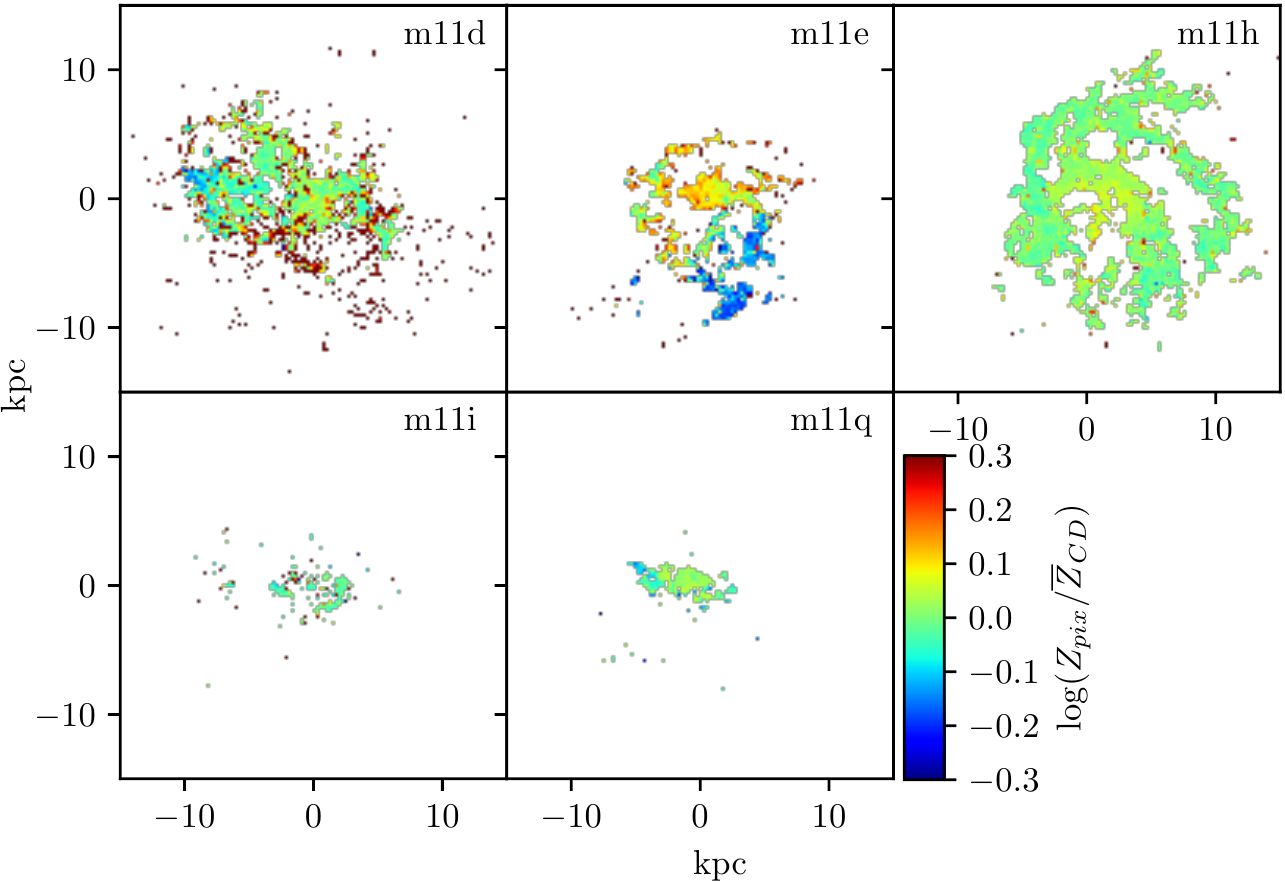}
    \caption{Spatial distributions of metallicity of cold and dense gas component (T$<500$K) of FIRE-2 galaxies with a pixel size of 250 pc, colored by their metallicity relative to solar, with the mean solar metallicity subtracted, at $t_{lookback}\approx$ 0.36 Gyr.
    \textit{Upper-left}: Galaxy \textbf{m11d} pre-starburst; LMC-mass galaxy that exhibits a majorly disrupted gas disk that is constantly expanding after multiple starbursts. Disk is well-mixed with regions of high enrichment.
    \textit{Top-center}: Galaxy \textbf{m11e}; main disk is well-mixed at $\log Z/Z_\odot \approx$-0.25 and exhibits an infalling satellite galaxy with a much lower metallicity ($\log Z/Z_\odot \approx$ -0.5.). The galaxies have not yet fully merged, with the satellite now near its first periapsis.
    \textit{Upper-right}: Galaxy \textbf{m11h}; LMC-mass small spiral galaxy with a well-mixed disk and a slight metallicity gradient.
    \textit{Lower-left}: Galaxy \textbf{m11i};  small galaxy near three times the SMC mass  with a disrupted gas disk from a recent starburst that is still relatively well-mixed. Metallicity is comparable to the main disk of m11e. 
    \textit{Bottom-center}: Galaxy \textbf{m11q}; another small galaxy just under twice the SMC mass undisturbed by starbursts or recent star formation.}
    \label{fig:allgals_imshows}
\end{figure*}

Figure~\ref{fig:allgals_imshows} displays the spatial metallicity distribution of the cold and dense gas for a single snapshot around $t_{lookback}\sim0.36$ Gyrs for all galaxies in this study at a pixel size of 250 pc. Colored by their difference from the mean metallicity of the individual galaxy (relative to solar), these maps allow us to characterize the galaxies at a glance in terms of their morphologies, apparent dynamical states, and sizes.

\begin{figure*}
	\includegraphics[width=\textwidth]{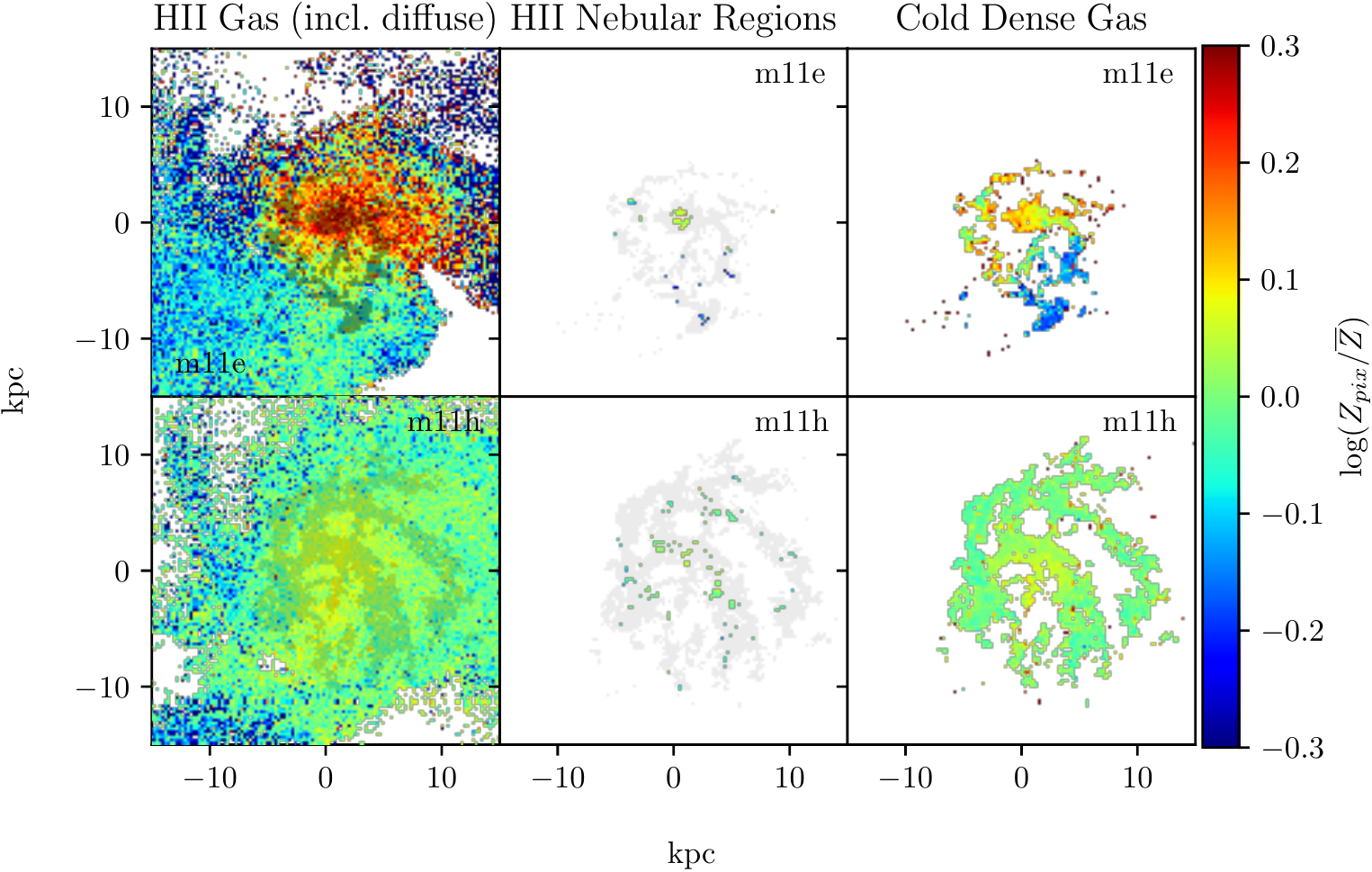}
    \caption{Metallicity distributions in various ISM components at 250 pc pixel size for two of the galaxies in this paper (m11e and m11h) from a single snapshot at $t_{lookback}\approx $0.36 Gyr, colored as Figure~\ref{fig:allgals_imshows}.
    \textit{Left column:} Hydrogen gas with $T \approx 10^4$ K (including diffuse ionized component). Top-left panel exhibits m11e's main body at a slightly higher metallicity envelope and the infalling companion. As the companion merges, the ionized gas halo is mixing into the main body. Bottom-left showcases the well-mixed hydrogen of m11h with a weak apparent gradient and some metal-poor gas mixing in the outskirts ($R \geq $10kpc).
    \textit{Middle column:} HII nebular regions ($T \approx 10^4$ K with $\sim $10 Myr star formation). Center-top panel shows a small distribution of gas in m11e with nebular regions tracing part of the infalling companion, and the center of the main body. Center-bottom shows how nebular emission tracks star-forming regions, closely tracing the dense gas peaks in m11h's spiral arm structure.
    \textit{Right column:} Cold and dense gas ($T < 500$ K) metallicity as shown in Figure~\ref{fig:allgals_imshows}, shown for comparison with diffuse HII gas and nebular regions.}
    \label{fig:m11e_m11h_imshows}
\end{figure*}

Figure~\ref{fig:m11e_m11h_imshows} displays the spatial metallicity distributions of the warm ionized phase (near the H$\alpha$ $T \approx 10^4$ K ridge-line) in two of the simulations in this study, \textbf{m11e} and \textbf{m11h}, in addition to including the before-seen cold and dense gas. We see immediately that the galaxies, as expected, are surrounded by a warm gas halo, and that ionized gas corresponding to HII/star-forming regions is dotted throughout the main body of the cold and dense gas in the galaxies. Below we further discuss the gas-phase metal distributions revealed in these two figures.

\subsection{Characterizing the Individual Simulations}

We find it important to briefly characterize each simulated galaxy (alphabetically), all originally introduced in \citet{El-Badry2018}, with the exception of m11q in \citet{Hopkins2018:fire}, to understand the idiosyncrasies that affect the metal distributions. For information on the stellar metallicity distributions, we direct the reader to \citet{Patel2021}.

An LMC-mass galaxy, \textbf{m11d} has a main body of cold and dense gas maintaining a metallicity within $\pm0.2$ dex of solar (fairly well-mixed throughout). The galaxy also displays numerous small pockets of metal-rich gas with metallicity nearly an order of magnitude larger than the primary body as a result of repeated feedback events that majorly disturb the gas. The snapshot of m11d in Figure~\ref{fig:allgals_imshows} represents the galaxy in a pre-major starburst state, an event we observe occurring $\sim$ 150 Myrs later that completely disrupts the main body of dense gas. Evident in Figure~\ref{fig:SFR_Mstar_overtime}, m11d experiences a significant drop in the star formation rate around this time. 

\textbf{M11e} displays a prominent main disk around $\sim$ 0.5 solar metallicity, as well as a second, metal-poor (metallicity nearly an order of magnitude below that of the main body) merging galaxy. The imaged snapshot occurs just as the cold and dense components of the satellite and primary body make contact, triggering several smaller feedback/starburst events in the main disk. Prior to merging, we note that each individual galaxy is reasonably well-mixed (within $\sim 0.1$ dex) in the cold and dense component of the ISM. 

Having the largest and most defined disk of the five, \textbf{m11h} exhibits an extremely well-mixed body (with a metallicity gradient of $\pm$0.1 dex) with few strong feedback events visible in the cold and dense gas. With a mass nearly identical to that of \textbf{m11d}, this galaxy also has a similar Gyr-averaged SFR, but remains the most consistent galaxy in terms of $<100$ Myr SFR, largely due to the lack of disruptive feedback taking place. A slight metallicity gradient is visible, to the discerning eye (see Figure~\ref{fig:CD_RvsTotalZ} for a quantitative presentation of the metallicity gradient).

 One of the smaller FIRE-2 dwarf galaxies in this analysis, \textbf{m11i} is around three times the mass of the SMC and exhibits several smaller feedback events. This is evident in the snapshot pictured in Figure~\ref{fig:allgals_imshows} through the small scatter of gas with metallicity over an order of magnitude higher than the rest. These starbursts remain infrequent, and this galaxy is similarly well-mixed like the three previously described. 

The lowest-mass and physically smallest galaxy in this study is \textbf{m11q} at just below twice the SMC mass. While similar to \textbf{m11i}, it has a  metallicity lower by $\sim$ 0.3 dex. We note that this galaxy's metallicity is nearly identical to that of the merging body in m11e, and is also strikingly well-mixed.

\subsubsection{Maps of $HII_Z$ vs $CD_Z$}

While we see most of the characteristics (such as disks) of our galaxies in the cold and dense gas component, it is important to analyze other components of the ISM, as the cold molecular gas is more difficult to observationally measure; more observable phases (such as the 'HII regions') may provide insight into how to interpret properties of the cold and dense gas phase. To provide a diverse example of these varying components, Figure~\ref{fig:m11e_m11h_imshows} displays two galaxies, \textbf{m11e} and \textbf{m11h}, in the three ISM components that we study in this paper: from left to right, we depict the HII gas (with diffuse component), nebular regions (HII gas with cospatial 10Myr SFR), and cold and dense gas (previously described in Figure~\ref{fig:allgals_imshows}). 

In the case of \textbf{m11e}, we can clearly see the warm ionized gas inflow from the merger being mixed into the main disk, including how well-mixed the individual components are independent of each other. It is also evident that the ionized gas extends across a larger region than the cold and dense gas, and so the ionized gas between the two galaxies begins interacting far before the cold and dense gas counterpart. Here we also note the inflow in both ISM components appears to be of much lower metallicity than the main body.

Similar comments can be made about \textbf{m11h} despite its absence of a merger. The ionized gas component extends beyond the cold and dense gas, and exhibits a similar metallicity gradient. We can see evidence of metal-poor inflows on the outskirts of the ionized gas and small areas of higher metallicity spread throughout. 

Upon examination of the nebular regions in both galaxies in Figure~\ref{fig:m11e_m11h_imshows}, we see that they are a clear tracer of the cold and dense gas, despite the differences between the spatial distributions of the two. This finding is particularly significant, because while the cold and dense phase of the ISM is typically more difficult to measure in observations, the regions of ionized gas with recent star formation ($<10$ Myr) may be able to provide an accurate estimate of the metallicity of the cold and dense gas component.

\subsection{Mass-Metallicity Relation (MZR)} \label{mzr}

\begin{figure*}
	\includegraphics[width=\textwidth]{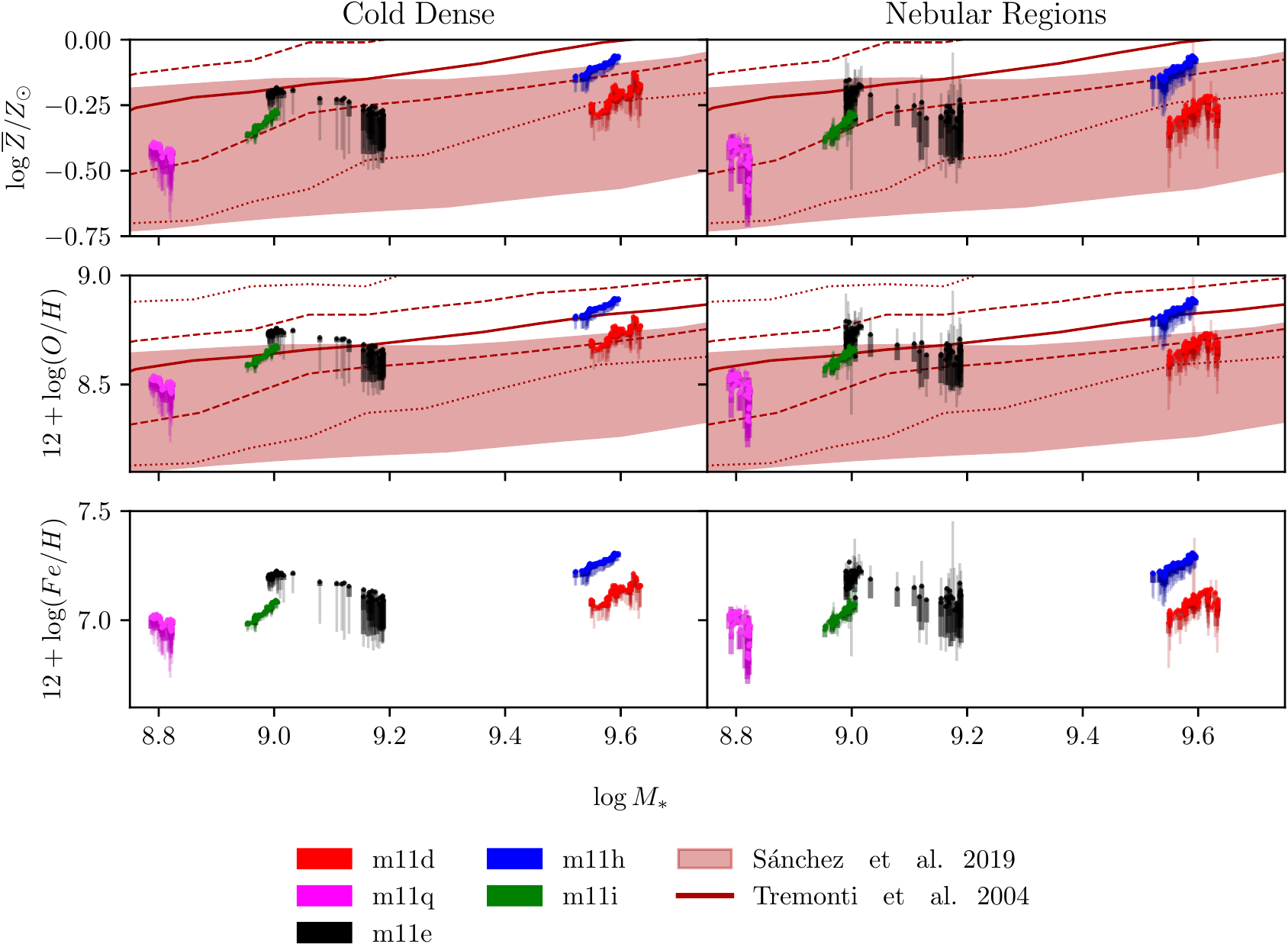}
    \caption{Gas-phase mass-metallicity relation (MZR) plots of five FIRE-2 dwarf galaxies in cold and dense gas ($T < 500$ K) and HII gas ($T \approx10^4$ K with recent SFR), in total metallicity (top row), O (center row), and Fe (bottom row). Each point represents different snapshots, denoting the average metallicity and stellar mass of all the designated ISM component gas in each, with colors as Figure~\ref{fig:SFR_Mstar_overtime}. Thick bars denote $25\%-75\%$ range of distribution of 250 pc pixels within each snapshot and thin bars represent $5\%-95\%$ range. Shaded red regions represent the range of observational data from \citet{Sanchez2019}. Dark red lines show observational data T04 from \citet{Tremonti2004}, with the solid line showing the median data points, the dashed lines showing -68\% and +68\%, and dotted lines as -95\% and +95\%. There seems to be little difference between cold dense and nebular regions, with nebular regions exhibiting scatter to slightly higher abundances in the tails of the distributions. Simulations are generally in agreement with estimated range of metallicities from \citet{Sanchez2019}. and \citet{Tremonti2004}.}
    \label{fig:mzr}
\end{figure*}

After exploring the spatial metallicity distributions of our selection, we investigate the correlation between gas-phase metallicity and stellar mass (MZR) in the range of masses, $M_* \sim 10^{8.75} - 10^{9.75} M_\odot$, in this study. Figure~\ref{fig:mzr} displays the gas-phase MZR in our selection of galaxies, averaged over all of the pixels from each snapshot, in both the cold and dense gas and the nebular regions ($T\approx10^4$K with $\sim$10 Myr star formation) for the total metallicity in solar units, O in 12+log(O/H) units, and Fe in 12+log(Fe/H) units, the latter of which is not typically observed in the gas-phase. The data from each simulation is plotted against observations from \citet{Sanchez2019} and \citet{Tremonti2004} in 12+log(O/H), which we scale to the appropriate units for the top panel. Because the gas-phase MZR varies depending on calibration, we note that this may cause a slight difference from the actual total solar MZR, but remains a good approximation for comparison.  

Most of the FIRE-2 dwarf galaxies that we analyze in this paper appear to match the observed MZR, with galaxies m11d, m11h, and m11i most closely following the observational distributions. M11h appears to fall just above the upper threshold of observations from \citet{Sanchez2019}. While m11e and m11q still fall within the range of observations, both exhibit a decreasing relationship of metallicity with stellar mass in Figure~\ref{fig:mzr} resulting in contrast to m11d, m11h, and m11i. 

Most of the galaxies in our sample (m11d, m11h, and m11i) are also in good agreement results found in other literature using FIRE to evaluate the MZR, including \citet{Ma2016} and \citet{Escala2018}, the former of which also compares FIRE galaxies in the MZR with observational data from \citet{Tremonti2004} at $z \approx 0$. 

There appears to be little difference between the two plotted ISM phases. One difference is that the warm ionized gas with recent star formation appears to have slightly higher scatter in the tails of the distribution, while the cold and dense gas displays less variation. Average metallicity appears to be identical with the exception of m11q and the final few snapshots of m11d, which exhibit a lower metallicity in the nebular regions than the cold and dense gas.

We note that m11e's metallicity experiences a sharp drop in the short timespan of two snapshots, approximately 50 Myrs. This is a rapid timeframe for a drop in metallicity of half a magnitude, and can be attributed to the major-merger at the time. \citet{Hopkins2013d} supports this, noting that major-mergers are known to cause such dramatic drops in gas-phase metallicity.

M11q, however, has no such apparent merger in our analyzed timeframe to explain such a significant drop in metallicity, and we see that this occurs throughout the entire span of the $\sim$1.4 Gyrs analyzed. Instead, this galaxy appears to be undergoing accretion of metal-poor gas onto its outskirts, explaining the steep negative metallicity profile.

We also note that each galaxy appears to have variations both within themselves at a certain time (a single point in Figure~\ref{fig:mzr}) and over time. For example, galaxies such as m11d, m11h, and m11i appear to be primarily consistent with their evolution, with m11d exhibiting more dependence on the scatter caused by major starbursts at later times. We can see that m11e possibly begins an evolution-dominated path before the satellite appears, resulting in much more variations within the "galaxy" itself and over time. Galaxy m11q, however, clearly has a distribution opposite that of the remaining galaxies, where the feedback caused by accretion prevents a clear evolution from taking place as variations inside the galaxy dominate over time.

\subsection{Metallicity Profiles} 
\begin{figure*}
	\includegraphics[width=\textwidth]{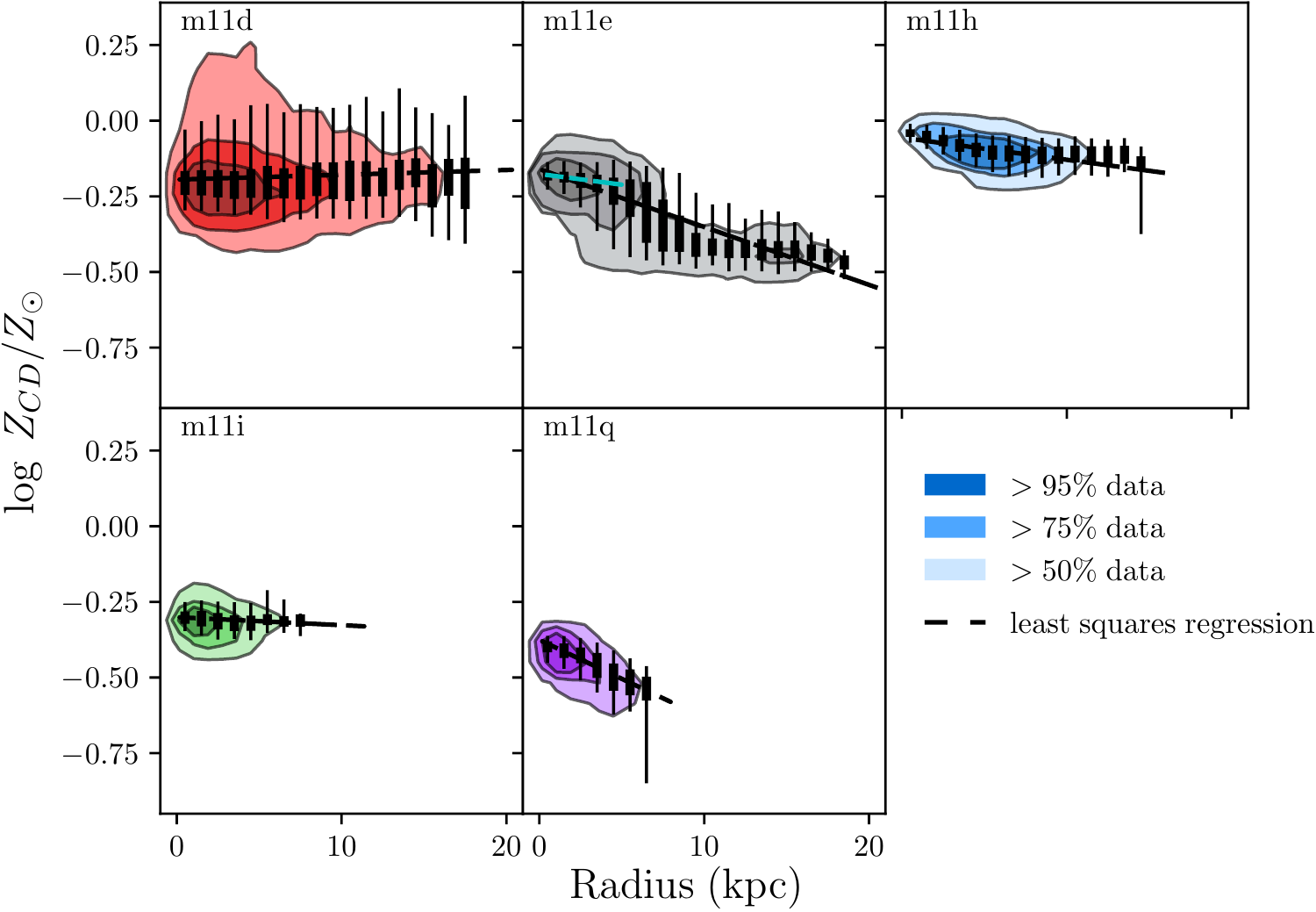}
    \caption{Radial profiles of the cold and dense gas metallicity in all five galaxies, analyzed here with 250 pc pixel size, where each distribution contains all pixels of the corresponding galaxy (across all snapshots). Shaded regions represent $50\%, 75\%$, and $95\%$ of data. Colors are as Figure~\ref{fig:SFR_Mstar_overtime}, with box and whiskers representing the same ranges as Figure~\ref{fig:mzr} in each radial bin. The least squares regression, denoted by the dashed black line, is fit through the entire distribution with the exception of m11e, where an additional cyan line has been plotted through only the main galaxy body before the satellite appears in the simulation (R$\geq$5 kpc and $t_{lookback}>$ 0.6 Gyr).
    \textit{Upper-left}: m11d; The galaxy is majorly disrupted due to starbursts, producing a positive metallicity gradient.
    \textit{Center-top}: m11e; An apparent steep metallicity gradient appears due to the merging companion. See cyan line for more accurate fit of main galaxy body (and further discussion in \S \ref{merging}).
    \textit{Upper-right}: m11h; An LMC-mass spiral with inside-out growth; slight negative metallicity gradient consistent with inside-out disk growth models.
    \textit{Lower-left}: m11i; Flat to slight gradient owing to a very well-mixed nature.
    \textit{Center-bottom}: m11q; A very steep gradient possibly due to metal-poor inflows that significantly affect the galaxy due to its small size and mass.}
    \label{fig:CD_RvsTotalZ}
\end{figure*}

We plot radial metallicity profiles of the FIRE-2 dwarf galaxies in the cold and dense gas in Figure~\ref{fig:CD_RvsTotalZ}. Typically, we would expect very slight negative relationships in the radial profile for dwarf galaxies, if they exist at all. \citet{Ma2017} notes that galaxies with strong perturbations usually have flatter gradients. These forms of feedback mix the gas, leading to metal distributions that are more uniformly distributed throughout the ISM than galaxies without as much feedback. Further, more massive galaxies with sufficient star formation and accretion will therefore have a negative gradient as they become more resistant to feedback, as supported by \citet{Bellardini2021}. Therefore, it is logical to conclude that dwarf galaxies, which are inherently more susceptible to feedback due to their small nature, will be dominated by perturbations and therefore exhibit extremely weak (if not flat) gradients \citep{El-Badry2016, Mercado2021}.

However, of the galaxies in this study, we note that not all follow these expectations, and we therefore find it necessary to iterate through the idiosyncrasies of the radial gradient of each galaxy.

M11d is the only galaxy in our selected sample that appears to exhibit a positive gradient, although it is slight. This galaxy also exhibits the largest spread in metallicity, covering just more than an order of magnitude. This galaxy undergoes several significant large-scale starburst/feedback episodes, the most of any galaxy studied here, so we consider this as a possible cause behind the positive gradient.

M11e is a particularly interesting case in terms of the metallicity profile. The merging body is clearly visible as a second distribution, and the significant drop in metallicity creates a particularly steep gradient. Because we recognize that the major-merger is the cause of this unique gradient, we chose to fit a line through the main body of m11e by limiting the radius to only the primary disk, and selecting to fit the corresponding line to snapshots where the satellite galaxy is not present in the cold and dense gas. As a result, we believe we obtain a far more realistic value for m11e's metallicity gradient in the cold and dense gas before the merger is present, represented by the dashed teal line, and we are therefore able to analyze the galaxy's gradient both pre-merger and during the merger, matching findings by \citet{Hopkins2013d}. We note here that our analysis timeframe concludes before the two bodies are able to effectively mix, which is why the gradient appears to be particularly steep. We predict that over time as the two galaxies begin mixing, the gradient will continue to flatten.

While it maintains a metallicity slightly higher than that of the pre-merger m11e, the body of m11h has a nearly identical metallicity gradient as the main body of m11e. Here, we see a smaller distribution than in the previous two galaxies, both in terms of size and spread. While this galaxy is smaller, this negative gradient matches what is expected for inside-out galaxy formation in disk galaxies of higher masses, both at observations \citep{Magrini2016, Sharda2021} and simulations \citep{Pilkington2012, Ma2017, Bellardini2021}. 

As seen in Figure~\ref{fig:allgals_imshows}, m11i is our smallest galaxy in the sample and displays the least scatter in its metallicity profile. A large majority of the gas is concentrated within 5 kpc and the metallicity stays within approximately 0.25 dex. This is the flattest gradient of our sample, and displays an extremely slight negative trend. 

M11q is similar to both m11h and m11i in terms of having little spread in metallicity over most of the galaxy, but it has a strong negative gradient. This gradient in particular is similar to that found in m11e post-merger, but there is no major-merger in this timeframe to explain such a stunningly steep gradient. In this case, we attribute this steep gradient to a high rate of metal-poor gas accretion in the galactic outskirts; the cold and dense gas reservoir appears to increase by roughly 5$\%$ over 200 Myr, and corresponds to the period where m11q's variations in metallicity are at their highest.

Similar gradients can be found in other literature on FIRE galaxies, such as in \citet{Ma2017}, despite the fact that they use high-redshift and higher-mass galaxies. 

\subsubsection{Mass-Metallicity Gradient Relation (MZGR)}

\begin{figure}
	\includegraphics[width=\columnwidth]{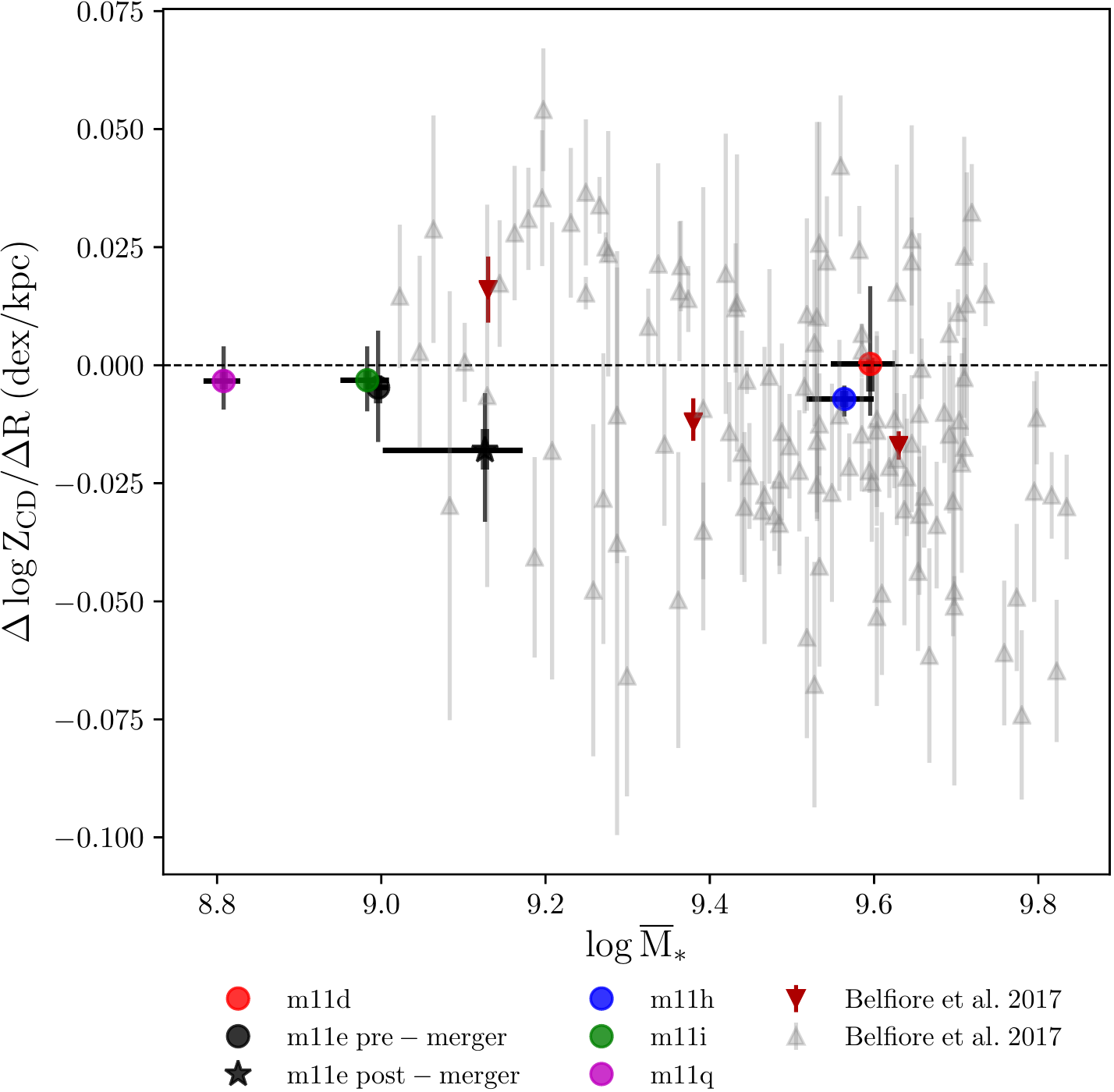}
    \caption{Cold and dense gas metallicity gradients as a function of stellar mass (MZGR) for all FIRE-2 dwarfs, with m11e being split into pre-merger (circle) and post-merger (star). Points denote the position of the galaxy's average stellar mass and average slope of metallicity gradient in dex per kpc (displayed in Figure~\ref{fig:CD_RvsTotalZ}). Thick vertical bars denote $25\%-75\%$  of data within that interval and thin vertical bars represent $5\%-95\%$ of data. Thick horizontal bars represent the change in stellar mass, with the leftmost point being the first snapshot and the rightmost point depicting the last snapshot. Colors are as Figure~\ref{fig:SFR_Mstar_overtime}. Grey and burgundy points and errorbars are observational data in the MaNGA survey from \citet{Belfiore2017}, where grey points are individual galaxies and red points are the median values of 0.25 dex stellar mass bins. Galaxies have weak to negative gradients; they exhibit a positive slope shortly after starbursts, as in the case of m11d. m11e clearly has a gradient similar to that of m11h and m11i until the merger. Simulations are consistent with observations. 
    }
    \label{fig:CD_slope_vs_stellarmass}
\end{figure}

We compare the MZGR in these simulations to that of data from the SDSS MaNGA Survey in \citet{Belfiore2017}, referenced in Figure~\ref{fig:CD_slope_vs_stellarmass}. However, we note to the reader that these observations are complete for stellar masses greater than $ 10^9 M_*$, and not all of our analyzed galaxies fall within this range. We plot all galaxies studied on this figure, splitting m11e into pre and post-merger in order to more accurately compare the galaxy's `true' gradient with and without the presence of the merging companion. For the galaxies that overlap with the mass range in \citet{Belfiore2017}, we see that the simulations are consistent with observations. While the scatter in individual galaxies of the observations appears to be far greater than our simulations, the median observational values are still consistent with our results.

Most gradients are flat or extremely close to zero, with m11d exhibiting a slightly positive scatter, despite having the same stellar mass as m11h. In addition, m11h exhibits the least amount of variation of its gradient. The three galaxies m11e (pre-merger), m11i, and m11q all have a roughly identical gradient, despite m11q's smaller stellar mass. We also make note of the significant drop in metallicity and increase in stellar mass for the separately distinguished post-merger phase of m11e shown in Figure~\ref{fig:CD_slope_vs_stellarmass}.

We include a version of Figures~\ref{fig:CD_RvsTotalZ} and \ref{fig:CD_slope_vs_stellarmass} using nebular gas metallicity in \ref{sec:HIIappendix}. The results are qualitatively similar, reassuring for our comparison with observations.

\subsection{Enrichment Variation Between ISM Phases}

Studying the relative metallicity enrichment between various phases of the ISM, such as the cold and dense gas and ionized gas, allows us to gain understanding as to whether the metals are moving spatially or between phases. All galaxies studied here appear to have a tight correlation in relative enrichment between the cold and dense and nebular ISM phases, with scatter remaining of approximately $\pm 0.1$ dex. 

\begin{figure*}
	\includegraphics[width=\textwidth]{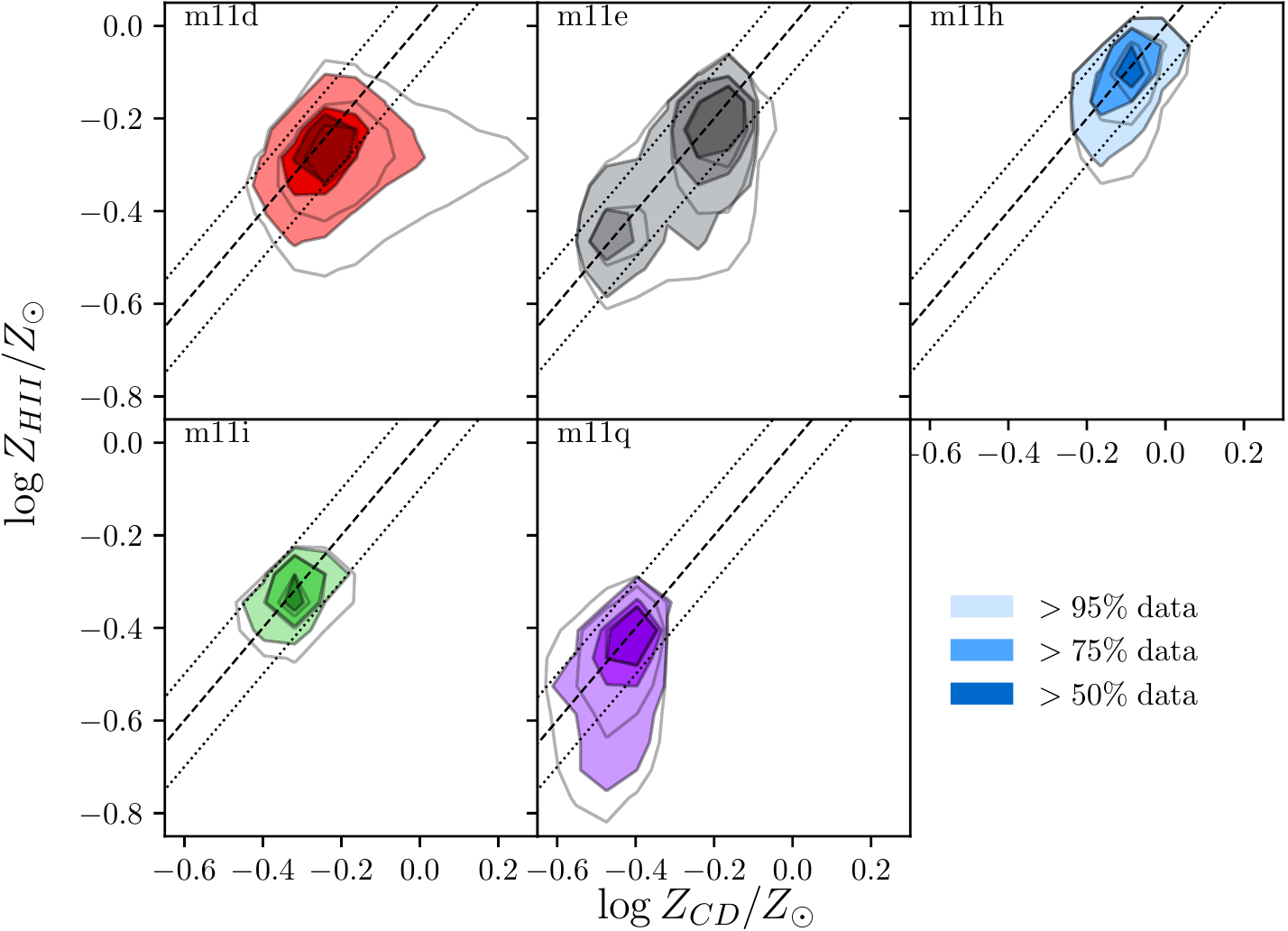}
    \caption{Comparing the enrichment of the cold and dense gas and HII gas (nebular and nebular+diffuse regions) in the five galaxies at 250 pc pixel scale for all snapshots. Filled contours denote $50\%, 75\%$, and $95\%$ of data inclusion in nebular regions; unfilled contours represent the nebular regions as well as diffuse HII component. Dashed line denotes equal abundances in the two phases, and dotted lines represent $\pm 0.1$ dex. Colors are as Figure~\ref{fig:SFR_Mstar_overtime}. In all cases, observable (nebular) regions are close to equal in abundance (with about $\pm 0.1$ dex scatter).
    Despite larger starbursts, m11d is mixed fairly well; but a diffuse HII component has a spray to high relative enrichment of the cold phase/low enrichment of the HII phase, which occurs at low cold dense surface density ($\Sigma_{CD} < 3 M_{\odot} {\rm pc}^{-2}$).
    m11e has a well-mixed distribution, but two clear groupings are visible between the galaxy and a companion. 
    m11h and m11i are both very well-mixed with no significant departures and little difference between nebular and diffuse HII. 
    m11q has a spray to low HII enrichment relative to the cold dense component, but is still relatively well-mixed.}
    \label{fig:contoursCDZvsHIIZ}
\end{figure*}

Figure~\ref{fig:contoursCDZvsHIIZ} provides an overview of the metallicity enrichment between the cold and dense gas and ionized gas. Remarkably, there is no significant bias in the average enrichment of phases in any of the galaxies. This occurs for both the diffuse and nebular HII, although we see slightly more scatter in the diffuse gas, which is consistent with other FIRE simulations at larger masses, such as MW mass galaxies in \citet{Bellardini2021}.

The two distributions in metallicity enrichment for m11e is particularly noticeable. This brings to attention that throughout m11e's merging process with its satellite galaxy, the two ISM phases of the merger \textit{still} primarily stay within $\pm0.1$ dex of each other. We conduct further investigation in \S \ref{merging}.

Also evident is m11q's steep metallicity scatter to low $Z$ in the nebular ISM. Again, this is a rather interesting result since m11q is one of the smallest galaxies in the sample, and it appears to exhibit as much (if not more) scatter as m11d, which also has significantly more feedback events that also occur on a much larger scale in relation to the size of the galaxy.

\begin{figure*}
	\includegraphics[width=\textwidth]{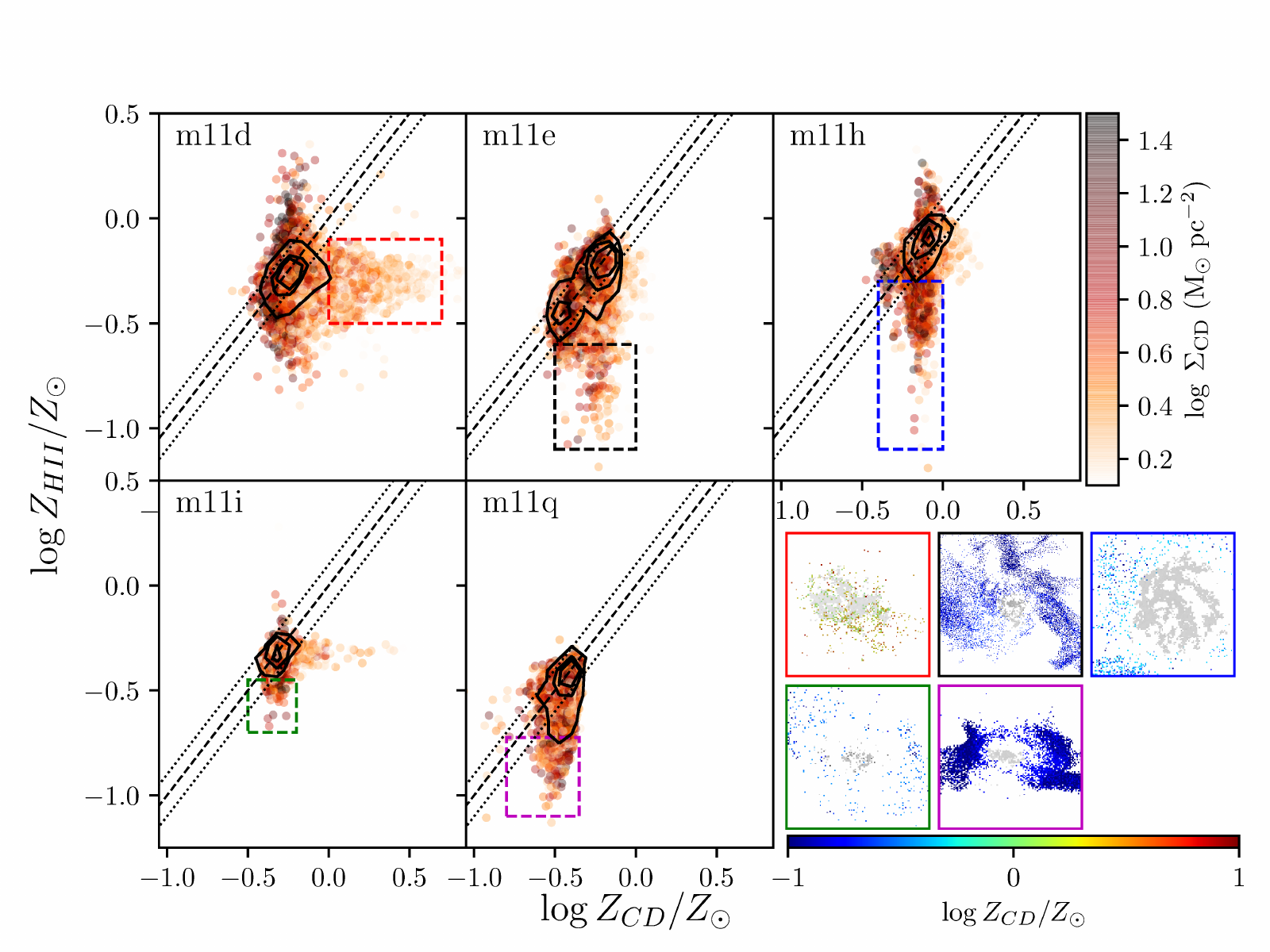}
    \caption{\textit{Larger five panels:} Comparison of  cold and dense gas abundances to HII gas abundances from all snapshots for each galaxy. Points are representative of every 250 pc square pixel, including diffuse HII gas, and are colored by the dex of the cold dense gas surface density ($M_{\odot} pc^{-2}$). Unfilled contours are the same filled contours (observable nebular regions with no diffuse gas) from Figure~\ref{fig:contoursCDZvsHIIZ}. Dashed and dotted lines are as Figure~\ref{fig:contoursCDZvsHIIZ}. Dashed rectangles, with colors as Figure~\ref{fig:SFR_Mstar_overtime},  represent areas of enrichment scatter from the one-to-one line where the cold and dense gas and HII gas are equally enriched. \textit{Smaller five panels:} In the lower-right corner, the color-coordinated image for each galaxy (ordered as larger panels) shows the spatial distribution of the boxed points in the larger five panels, highlighting their origin. The full $Z_{CD}$ distribution in grey, similar to those shown in Figure~\ref{fig:m11e_m11h_imshows}. M11d's imaged scatter in the red panel, where $Z_{CD}/Z_{\odot} > 0$, clearly demonstrates that scatter in the $+$x direction is resultant of metal-rich starbursts. The analyzed scatter for m11e , m11h, m11i, and m11q all show metal-poor HII gas falling into the galaxies, and in cases like m11q, these inflows contribute to a particularly steep metallicity gradient.}
    \label{fig:scatterCDZvsHIIZ}
\end{figure*}

In order to understand the cause of the scatter outside $\pm0.1$ dex, we investigate the outlying pixels in a single snapshot. Figure~\ref{fig:scatterCDZvsHIIZ}'s smaller five panels in the lower-right represent the snapshot at $t_{lookback}=0.36$ Gyrs. This figure shows the metallicity relative between the cold and dense gas and ionized gas for every pixel, with the same filled contours as Figure~\ref{fig:contoursCDZvsHIIZ} for reference, but we study several regions of outlier pixels individually.

The scatter seen occurring in the cold and dense gas (at constant $Z_{\rm HII}$) appears to primarily consist of pixels that have a low surface density in the cold and dense component. We also note that scatter predominately seems to occur at either constant $Z_{\rm CD}$ or $Z_{\rm HII}$; we see very little deviation outside of relative pure $Z_{\rm CD}$ or $Z_{\rm HII}$ enrichment . 

To understand this finding in Figure~\ref{fig:scatterCDZvsHIIZ}, we supplement a visual for the physical processes of the scatter in the form of a panel of insets showing the locations of pixels in the boxed regions of metallicity space. Only the outlying pixels are colored, which allows us to easily identify exactly which pixels are the culprit. There is a clear connection in what seems to be causing the directional scatter. From these inset images, we generally conclude here that positive scatter in the cold and dense gas is caused by metal-rich starbursts, and negative scatter in the ionized gas is caused by metal-poor inflows into the galaxy. 

Reaching back to the peculiar case of m11q, it can easily be seen how the metallicity profile is so influenced. As discussed earlier, dwarf galaxies are ideal targets for studying the effects of feedback since it is more well-pronounced in such a small body. m11q is both the smallest galaxy in the sample, and appears to have a large amount of extremely metal-poor gas accretion. This can directly explain the results of Figure~\ref{fig:CD_RvsTotalZ}, and why the gradient of m11q is far more influenced by this accretion.

\subsection{Merging Metallicity Distributions} \label{merging}

M11e is unique in that it has a clear major merger during the t$\sim$1.4 Gyrs studied. We find it important to further analyze the effect of the merger on the primary galaxy in terms of metallicity enrichment patterns. 

\begin{figure*}
	\includegraphics[width=\textwidth]{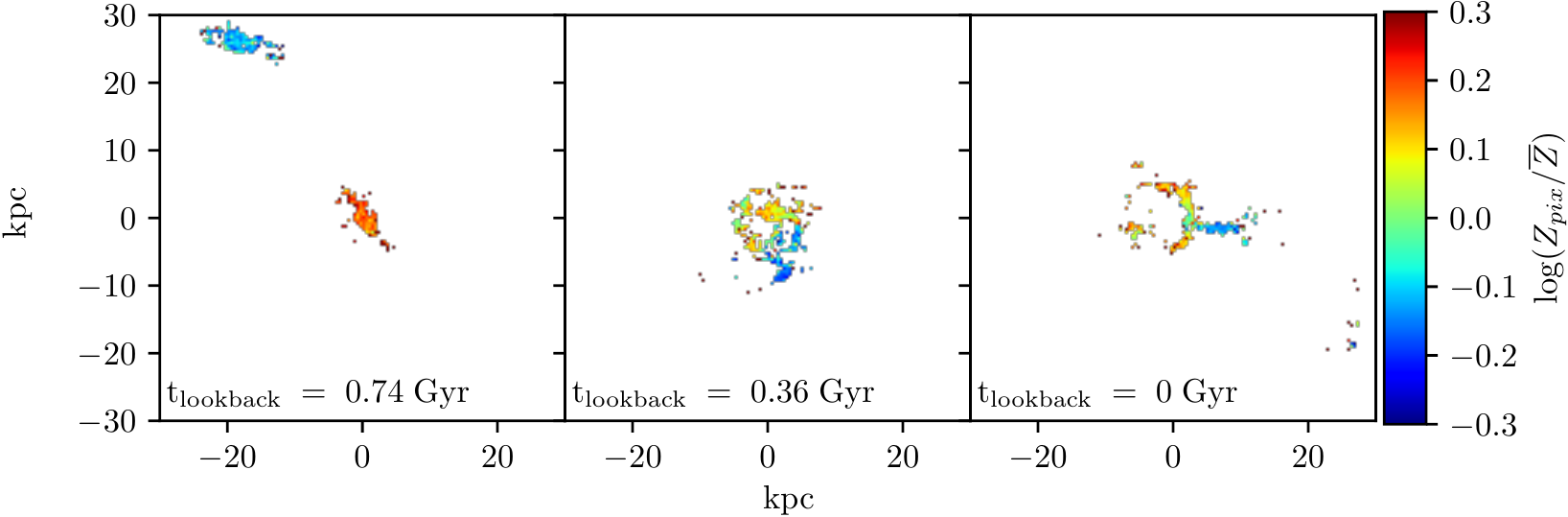}
    \caption{Spatial distributions of cold and dense gas ($T < 500$ K of m11e with a pixel size of 250 pc in three snapshots ($t_{lookback}\approx$ 0.76, 0.36, 0 Gyrs, respectively), colored as Figure~\ref{fig:allgals_imshows}. 
    \textit{Left}: One of the first appearances of the merging companion galaxy in our analysis. On first infall, the cold components have not yet mixed.
    \textit{Center}: Companion galaxy is close to pericenter, and the cold components have begun to mix.
    \textit{Right}: Merging galaxy causes a significant starburst in the main disk of m11e, producing a tennis racquet shape, with the "head" being the shell of the main disk (starburst having occurred in the center) and the satellite galaxy forming the "handle" in this projection. }
    \label{fig:m11e_imshows}
\end{figure*}

Figure~\ref{fig:m11e_imshows} shows the merger in the cold and dense gas, from roughly the time the companion appears in our analysis ($t_{lookback}=0.76$ Gyrs) to the end of our analysis period ($t_{lookback}=0$ Gyrs). We see that the infalling galaxy appears to be roughly the same size as the main body, but with a metallicity that is nearly half a dex lower. Both galaxies appear to have their metallicity nearly uniformly distributed in the cold and dense gas.

\begin{figure}
	\includegraphics[width=\columnwidth]{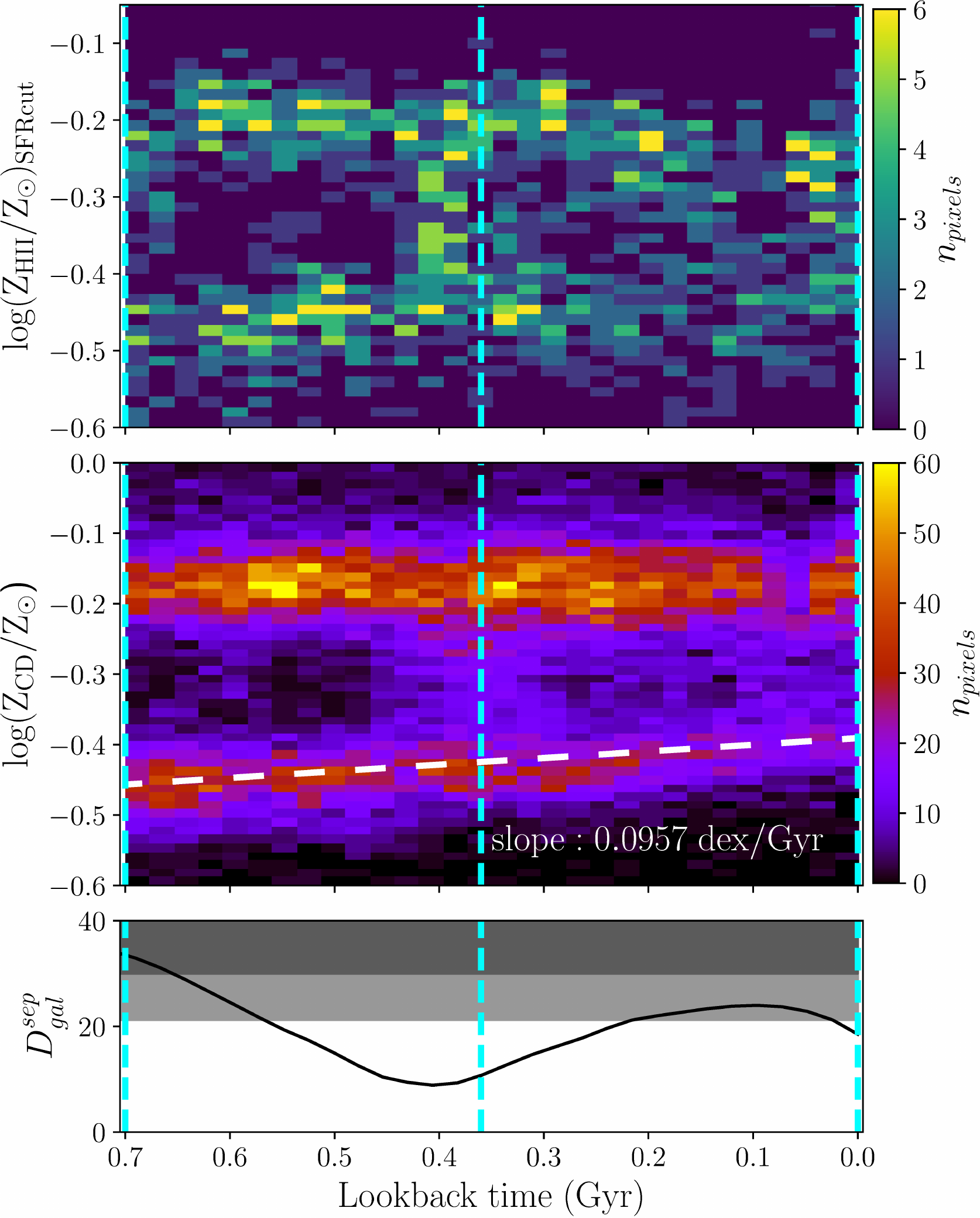}
    \caption{Metallicity distribution in nebular regions ($T\approx10^4 K$) (top panel) and metallicity distribution in cold dense gas ($T < 500$ K) as a function of time (center panel). Bottom panel displays separation distance between the main disk of m11e and the merging companion. Cyan lines denote the time of snapshots seen in Figure~\ref{fig:m11e_imshows}. The white dashed line represents a least squares regression of the satellite galaxy's metallicity distribution in the cold and dense gas, where an overall slope of approximately 0.1 dex/Gyr is observed. Both gas phases show a bimodal distribution resulting from the merger, beginning close to its first appearance in the cold dense gas at t$\sim$0.67 Gyr.}
    \label{fig:m11e_heatmaps}
\end{figure}

As we see time pass and the satellite spirals into into the main body, the cold and dense components appear to first make contact and complete the first periapsis around $t_{lookback}\sim0.36$ Gyrs, the second panel of Figure~\ref{fig:m11e_imshows}. The cold and dense phases of the two galaxies continue to swirl and interact, setting off minor feedback events, before finally setting off a major starburst that disrupts the entirety of m11e's main body in the last snapshot ($t_{lookback}=0$ Gyrs), serving as an interesting shape reminiscent of a tennis racquet. Over this same analysis period, we can see the satellite very slightly increase in its overall metallicity, which is better viewed in Figure~\ref{fig:m11e_heatmaps}. In addition, the primary body of m11e has a significant drop in the metallicity by the time the first periapsis occurs.

While previous literature such as \citet{Torrey2012} indicates that major mergers in disk galaxies typically flatten the metallicity gradient through metal-poor inflows, we would like to not that the merging of m11e, as a dwarf galaxy, is a different scenario. Representative through both Figure~\ref{fig:CD_RvsTotalZ}, \ref{fig:CD_slope_vs_stellarmass}, and \ref{fig:m11e_imshows}, the gradient of m11e does not flatten at all but steepens as a result. Therefore, the analysis of major mergers between larger disk galaxies and dwarf galaxies, like m11e, may result in different effects on the metallicity gradients.

While looking at the spatial maps of the cold and dense component is illuminating, we are equally interested in the metallicity as a function of time. We therefore further analyze the distributions of metallicity in both galaxies through both the nebular regions (HII with recent star formation) and the cold and dense gas in Figure~\ref{fig:m11e_heatmaps}. In addition to seeing the two distributions in each ISM component, the slope of the satellite's metallicity becomes apparent, particularly through a regression line fit through the cold and dense gas metallicity in the middle panel of Figure~\ref{fig:m11e_heatmaps}. While the galaxy appears to have an upward trend in metallicity, the fitted line's slope represents a slope of about 0.1 dex/Gyr in the cold and dense gas. We find that this slope is consistent with self-enrichment from internal star formation in the satellite's main body. 


\section{Discussion}

In this paper we have explored the stellar mass--gas-phase metallicity relation (MZR) and metallicity gradients in a suite of FIRE-2 dwarf galaxies. These relations connect the products of stellar evolution to the evolutionary history of the galaxy and have been studied for decades (e.g., \citealt{Vila-Costas1992}; \citealt{Tremonti2004}). 
Assuming a so-called closed-box model (i.e., no inflow or outflow of gas into the galaxy), the metal enrichment would be consistent with the integral of the star-formation rate over cosmic time (star formation history).  This would give a linear MZR and metal profiles that reflect higher metal content with galaxies/regions of higher star formation rate \citep{Ellison2008}.
Dwarf galaxies systems are far from closed-box. They are more susceptible to feedback events and therefore prone to outflows, and are more disrupted by mergers and starbursts. 
Our findings reflect this; in regards to the MZR, while we find a general linear increase in metallicity with stellar mass, we also find a large amount of scatter can exist among galaxies of similar masses, and that this scatter is consistent with observations \citep[][]{Sanchez2019}. Merging systems produce the most scatter in both stellar mass and in metallicity. We also find that the FIRE-2 dwarfs are consistent with observations of the median metallicity gradients observed by the MaNGA survey \citep[][]{Belfiore2017} and that most gradients are flat or slightly negative, although individual observed galaxies seem to exhibit more scatter. We show that mergers and metal poor gas inflows produce steeper gradients (e.g. Figure~\ref{fig:CD_RvsTotalZ}).
Our results therefore suggest that outflows, inflows,  as well as mergers play an important role in the overall metal-content and metal profile of galaxies and in particular are especially important for interpreting metallicity observations of dwarf galaxies \citep{Hopkins2013d}. 

We find that turbulent mixing is important for dwarf galaxies' metallicity profiles and a lack of observed correlation between metals and SFR (see Appendix \ref{Z_and_SFR}).  Starbursting events (e.g., m11d), mergers (e.g., m11e), and  spiral-like structure (e.g. LMC-like m11h) all produce mild negative or flat gradients, and show consistency with metal expectations from inside-out star formation \citep{Krumholz2018}. However, given the short sound-crossing time for dwarf galaxies, mixing is highly efficient and quickly acts to flatten metal gradients. While we note that regions of high star formation rate correspond to regions of enhanced metals and tighter correlation between metals found in cold gas and nebular gas  (see Figure~\ref{fig:contours_SFRfortyvsD}), efficient turbulent mixing quickly diminishes correlations between star formation and metal enhancements. Regions of moderate to low SFR have no correlation with metal content and more scatter exists between cold and nebular gas.   Indeed observational studies  have found weak or no correlation between SFR and gas-phase metallicity. The strongest correlations tends to be observed only in regions of high SFR \citep[][]{Lara-Lopez2010, Salim2014, Telford2016, Sanchez2019}. 

This remains consistent with previous work in the FIRE simulations. For example, related literature on MW-mass disk galaxies and their effect on gradients can be applied to m11h, such as \citet{Ma2017} and \citet{Bellardini2021}'s findings that rotationally-supported disky galaxies are more likely to have a negative gradient, which we see in Figure~\ref{fig:CD_RvsTotalZ}, and correlate with \citet{El-Badry2018}, where they classify m11h as having significant rotational support. Less perturbed galaxies are also found to have flat gradients, as visible in m11i, where fewer starbursting events occur, versus the differing gradients of m11d, m11e, and m11q, which are all highly perturbed by mergers, starbursts, or metal-poor diffuse gas inflow \citep{Ma2017, Bellardini2021, Patel2021}. 

This work further explores the relation between different ISM phases (cold, atomic/molecular, and dense vs. warm, ionized, and diffuse) and their metal content. Observationally, it is difficult to probe the cold molecular and atomic gas phase metallicity as the most common tracers are in the ionized phase (e.g., ionized oxygen).  Studying the relative abundance of metals in cold gas has important implications for mixing time scales and the metal content of the next generation of stars, since star formation occurs in cold molecular gas and not in ionized warm gas.   Overall, we find that all galaxies studied here appear to have a tight correlation in relative enrichment between the cold and dense and nebular ISM phases. Turbulent mixing provides a natural mechanism to quickly diffuse metals from one phase to another and the cascading rates measured in different ISM phases via power spectral analysis and other statistics are similar \citep[][]{Herron2016,Pingel2018, Burkhart2021}. This tight correlation breaks down in the outskirts of dwarf galaxies, where almost all of the gas content is ionized and the regions are subjected to metal poor inflows. Additionally, star bursting regions can also contribute to an enhancement of metals in the cold and dense phase relative to the ionized phase. Overall, for the bulk of the gas in the central regions of dwarf galaxies, metallicity estimates from observations of nebular regions  are strongly predictive of the cold and dense gas metallicity.

\section{Conclusions}

We presented an analysis of spatially-resolved gas-phase metallicity relations for five FIRE-2 dwarf galaxies with masses between the SMC and LMC in this paper. We investigated metallicity distributions across several different ISM phases, the gas-phase mass-metallicity relation, metallicity profiles and gradients, the relative enrichment between the ISM phases, and and enjoyed a brief case study of a major merger into a major-merger in one run (m11e). We can summarize our findings by the following key takeaways:
\begin{enumerate}
    \item All five FIRE-2 dwarf galaxies match the observable gas phase MZR. We see falling metallicities in time in m11e (explained by the merging metal-poor satellite galaxy) and m11q (reasoned through the accretion of metal-poor ionized gas). The three other galaxies, m11d, m11h, and m11i, all have positive relationships at a fairly consistent rate. 
    \item Metallicity in 'nebular regions' closely traces metals in the cold and dense gas component, resulting in a typical scatter of $\pm$0.1 dex. By extension of this, we conclude that metallicity estimates from observations of 'nebular regions' (ionized gas where there is recent star formation) are strongly predictive of the cold and dense gas metallicity. 
    \item The relative metallicity enrichment between the ionized gas and cold and dense gas reservoirs does not appear to be influenced by local SFR (see Figure~\ref{fig:scatter_SFRfortysdvsZ}). Instead, it can be linked to large-scale dynamics of the ISM, like gas inflows or galaxy-scale starbursts, as seen in Figure~\ref{fig:scatterCDZvsHIIZ}.
    \item All our simulated galaxies exhibit flat to slightly negative metallicity gradients, and are consistent with observations, albeit with slightly more scatter. Results in the ionized gas are identical to those in the cold and dense gas. Our lowest mass simulations exhibit a large scatter in their gradients over the analysis period due to their various dramatic disruptions to their gas dynamics.
    \item The merging companion in m11e results in an increased stellar mass, but a drop in average gas-phase metallicity as the metal-poor satellite begins to incorporate itself into the main body.
\end{enumerate}

\section*{Acknowledgements}

L.E.P. is grateful to the University of Louisville's Brown Fellows Program for financial support, funded by the James Graham Brown Foundation. BB is grateful for generous support by the David and Lucile Packard Foundation and Alfred P. Sloan Foundation. The Flatiron Institute is supported by the Simons Foundation. AW received support from: NSF via CAREER award AST-2045928 and grant AST-2107772; NASA ATP grant 80NSSC20K0513; HST grants AR-15809 and GO-15902 from STScI.

\section*{Data Availability}

The data supporting the plots within this article are available on reasonable request to the corresponding author. A public version of the GIZMO code is available at \url{http://www.tapir.caltech.edu/~phopkins/Site/GIZMO.html}. Additional data including simulation snapshots, initial conditions, and derived data products are available at \url{https://fire.northwestern.edu/data/}. The FIRE-2 simulations are publicly available \citep{Wetzel2022} at \url{http://flathub.flatironinstitute.org/fire}.



\bibliographystyle{mnras}
\bibliography{ref,library} 




\appendix

\section{Metallicity Profiles and MZGR in HII}
\label{sec:HIIappendix}

\begin{figure*}
	\includegraphics[width=\textwidth]{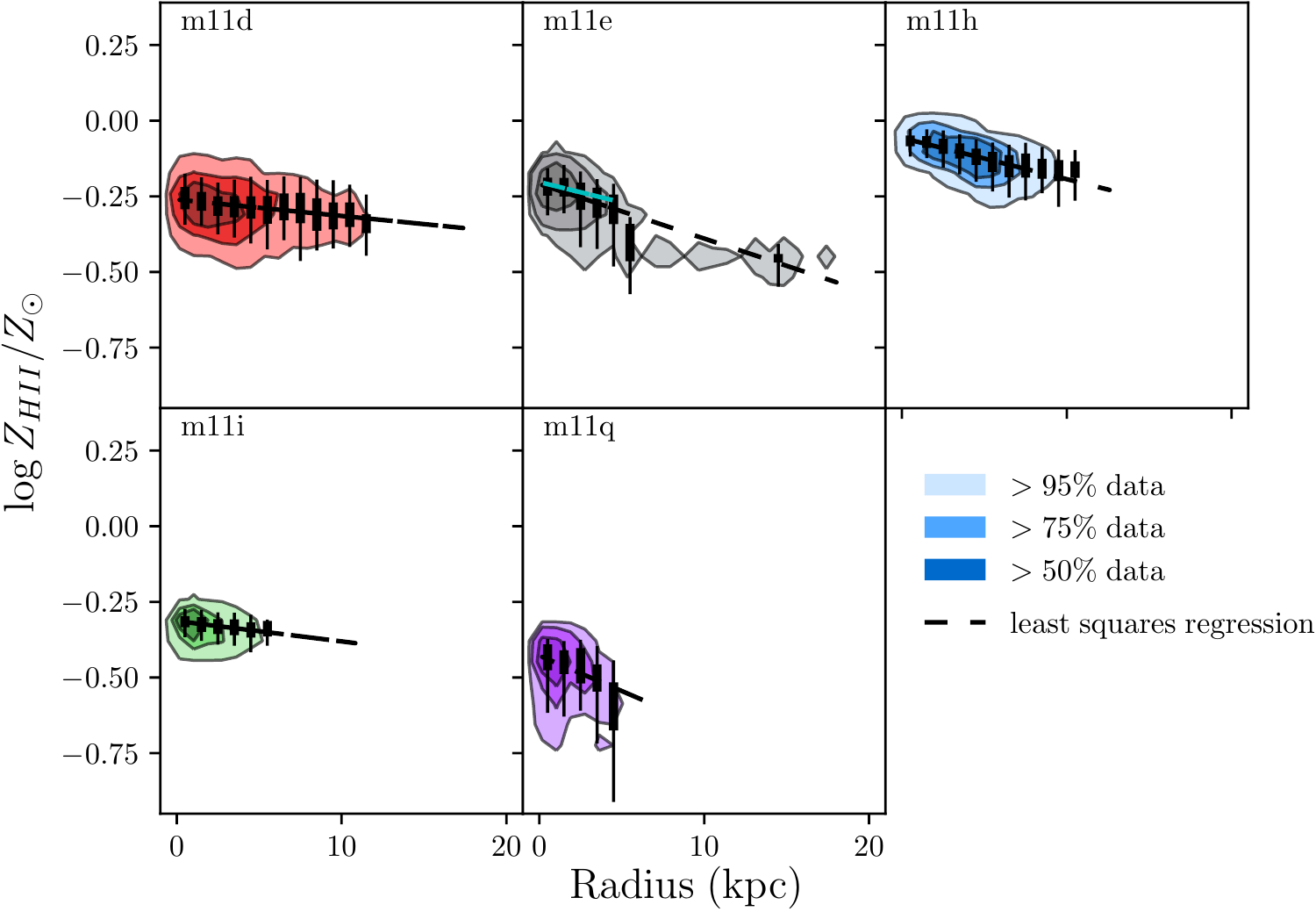}
    \caption{Metallicity profiles in the ionized gas (with recent star formation) of the five galaxies analyzed here with 250pc pixel size, where distribution contains all pixels of the corresponding galaxy across all snapshots. Shaded regions represent $50\%, 75\%$, and $95\%$ of data. Colors are as Figure~\ref{fig:SFR_Mstar_overtime} and box and whiskers are as Figure~\ref{fig:mzr}. The least squares regression, denoted by the dashed black line, is fit through the entire distribution with the exception of m11e, where an additional cyan line has been plotted through only the main galaxy body (R$>$5 kpc and t$<$ .83 Gyr). 
    \textit{Upper-left:} Galaxy m11d; majorly disrupted due to starbursts with a negative metallicity gradient.
    \textit{Center-top:} Galaxy m11e; apparent steep metallicity gradient due to the merging companion. See cyan line for fit of main galaxy body. Both fits appear identical due to the ionized gas regions extending much farther and therefore mixing much earlier than their cold and dense counterparts.
    \textit{Upper-right:} Galaxy m11h; appears to be an LMC-mass spiral with inside-out growth; slight metallicity gradient most consistent with such growth models.
    \textit{Lower-left:} Galaxy m11i; slight gradient owing to very well-mixed nature.
    \textit{Center-bottom:} Galaxy m11q; very steep gradient possible due to metal-poor inflows that significantly affect the galaxy due to its small size and mass.}
    \label{fig:HII_RvsTotalZ}
\end{figure*}

\begin{figure}
	\includegraphics[width=\columnwidth]{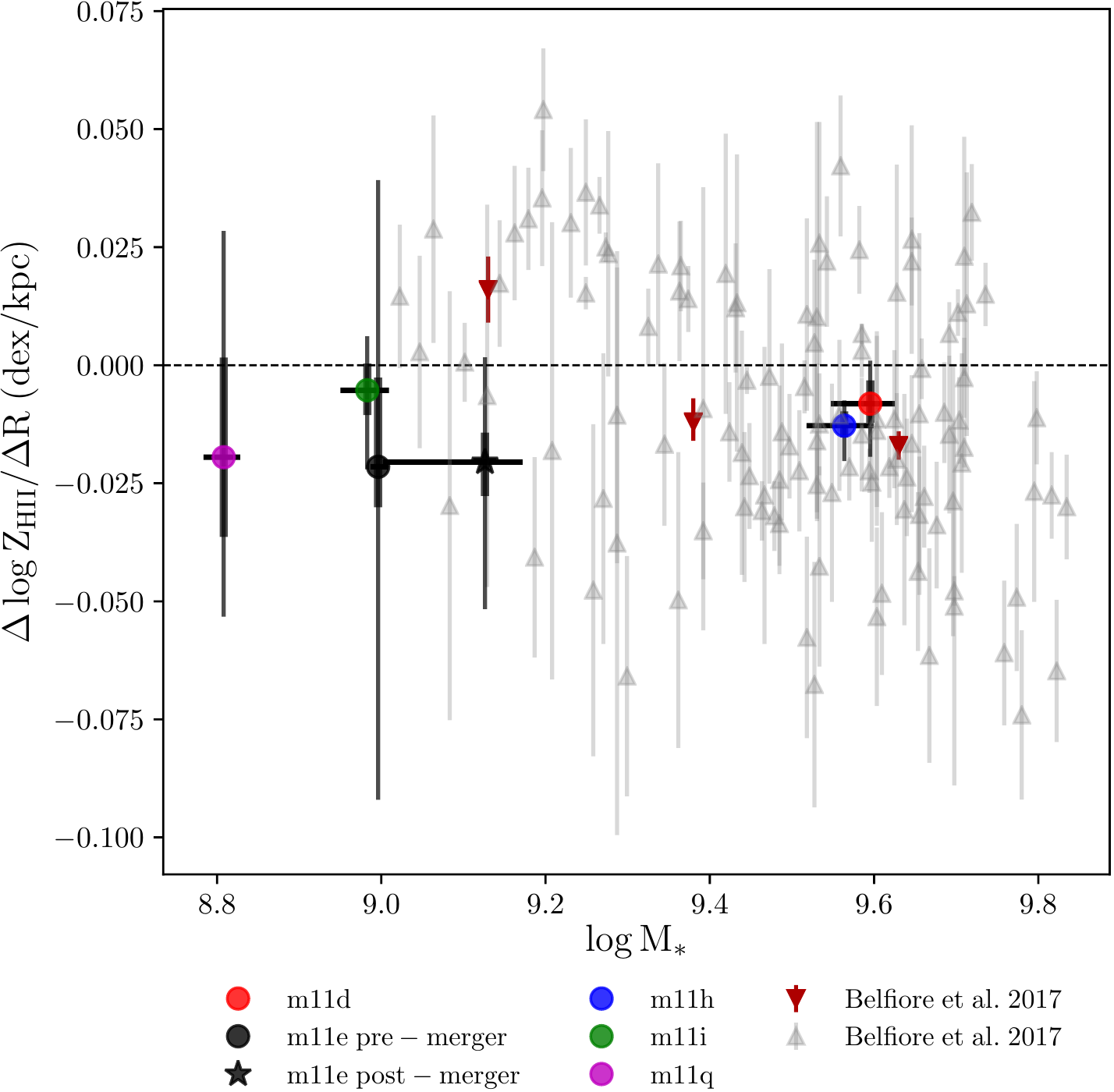}
    \caption{Ionized gas (with recent star formation) metallicity gradients as a function of stellar mass for all FIRE-2 dwarfs, with m11e being split into pre-merger (circle) and post-merger (star). Points denote the position of the galaxy's average stellar mass and average slope of metallicity gradient in dex per kpc (displayed in Figure~\ref{fig:CD_RvsTotalZ}). Thick vertical bars denote $25\%-75\%$  of data within that interval and thin vertical bars represent $5\%-95\%$ of data. Thick horizontal bars represent the change in stellar mass, with the leftmost point being the first snapshot and the rightmost point depicting the last snapshot. Colors are as Figure~\ref{fig:SFR_Mstar_overtime}. Grey points and burgundy points and errorbars are observational data in the MaNGA survey from \citet[][; their Figure B1]{Belfiore2017}, where grey points are individual galaxies and red points are the median values of 0.25 dex stellar mass bins. Galaxies primarily have a weak to negative gradient; they may exhibit a positive slope shortly after starbursts, as in the case of m11d. M11e clearly has a gradient similar to that of m11h and m11i before its merger.. Simulations are in good agreement with observations.  }
    \label{fig:HII_slope_vs_stellarmass}
\end{figure}

Here we present the metallicity profiles and gradients as a function of stellar mass (as in Figures~\ref{fig:CD_RvsTotalZ} and \ref{fig:CD_slope_vs_stellarmass}) for the nebular regions in Figures~\ref{fig:HII_RvsTotalZ} and \ref{fig:HII_slope_vs_stellarmass}. Though qualitatively similar, there is significantly more snapshot-to-snapshot scatter in the overall gradients in the lower mass galaxies (particularly m11e and m11q). This is largely due to the low SFRs (i.e., small number of pixels contributes to the gradient calculation) and the large flux of warm gas, either in the halo-halo interaction in the merger for m11e or the gas accretion for m11q.

\section{Differential Enrichment Distance (DED)}

 In Figure~\ref{fig:contoursCDZvsHIIZ} we found that the metallicity of the cold and dense gas and the ionized gas phases were nearly identical (within $\approx$0.1 dex). However, we also explore this by the Differential Enrichment Distance (DED). 

\begin{equation}
DED = \frac{\log Z_{CD}/Z_{\odot} - \log Z_{HII}/Z_{\odot}}{\sqrt{2}}
\end{equation}

A positive DED value indicates higher local enrichment in the cold and dense gas, while a negative DED value indicates higher local enrichment in the ionized gas. 

\begin{figure*}
	\includegraphics[width=\textwidth]{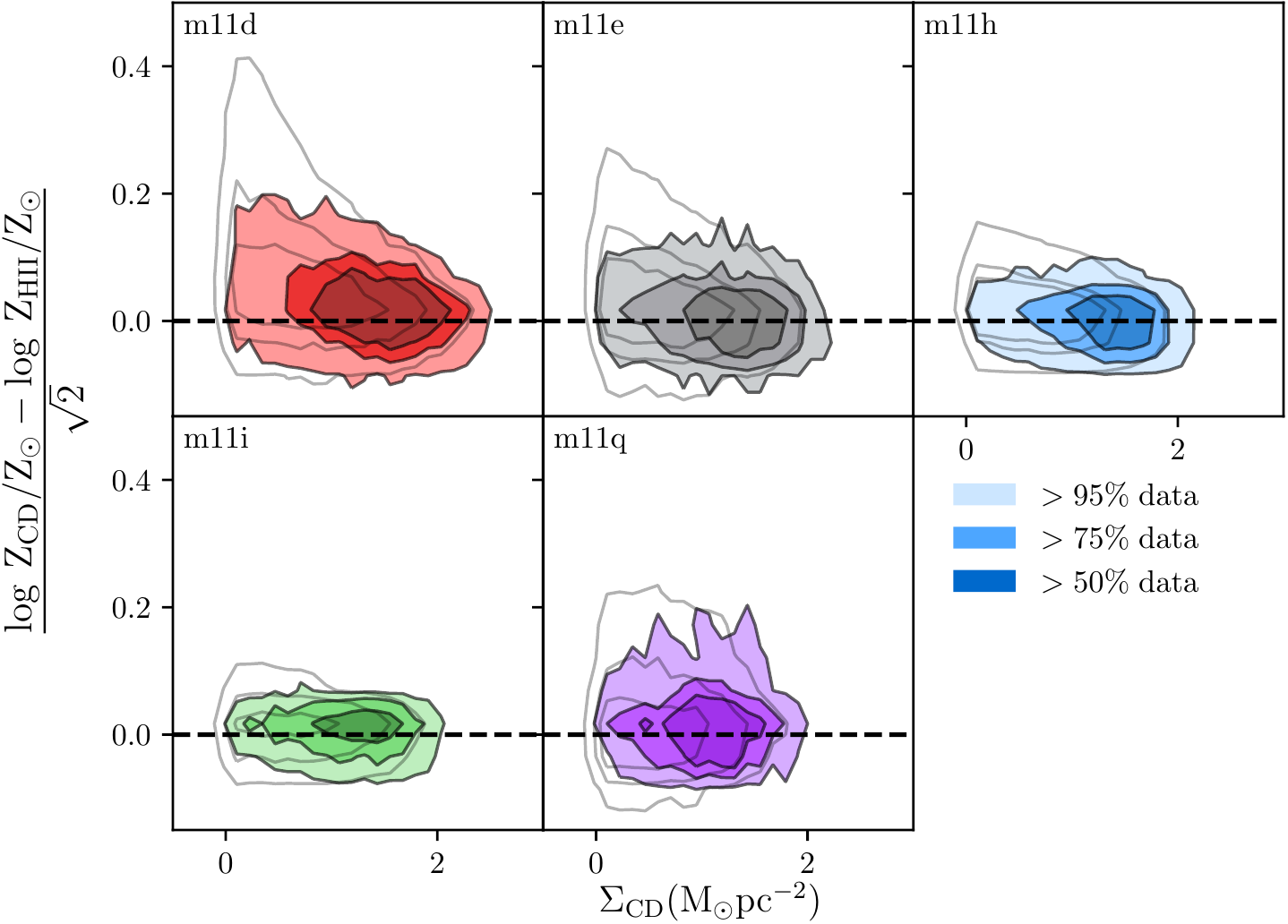}
    \caption{Comparing the distribution of cold and dense gas surface density to the relative Differential Enrichment Distance (DED) for all five FIRE-2 dwarf galaxies. Filled contours denote $50\%, 75\%$, and $95\%$ of data inclusion in nebular regions with no diffuse gas; unfilled contours represent the with the addition of the diffuse component. Dashed black line represents equal enrichment in both the HII gas and the cold dense gas. Colors are as Figure~\ref{fig:SFR_Mstar_overtime}. The filled contours (observable) appear to be roughly equally enriched, except for m11d and m11i, the two galaxies with the most starburst activity. The biggest scatters exist at low surface density and are not evident in the nebular regions. No general dependence on the differential enrichment as a function of cold and dense gas surface density is seen in any of the simulations.}
    \label{fig:contours_masscut_CDMassvsD}
\end{figure*}

\begin{figure*}
    \includegraphics[width=\textwidth]{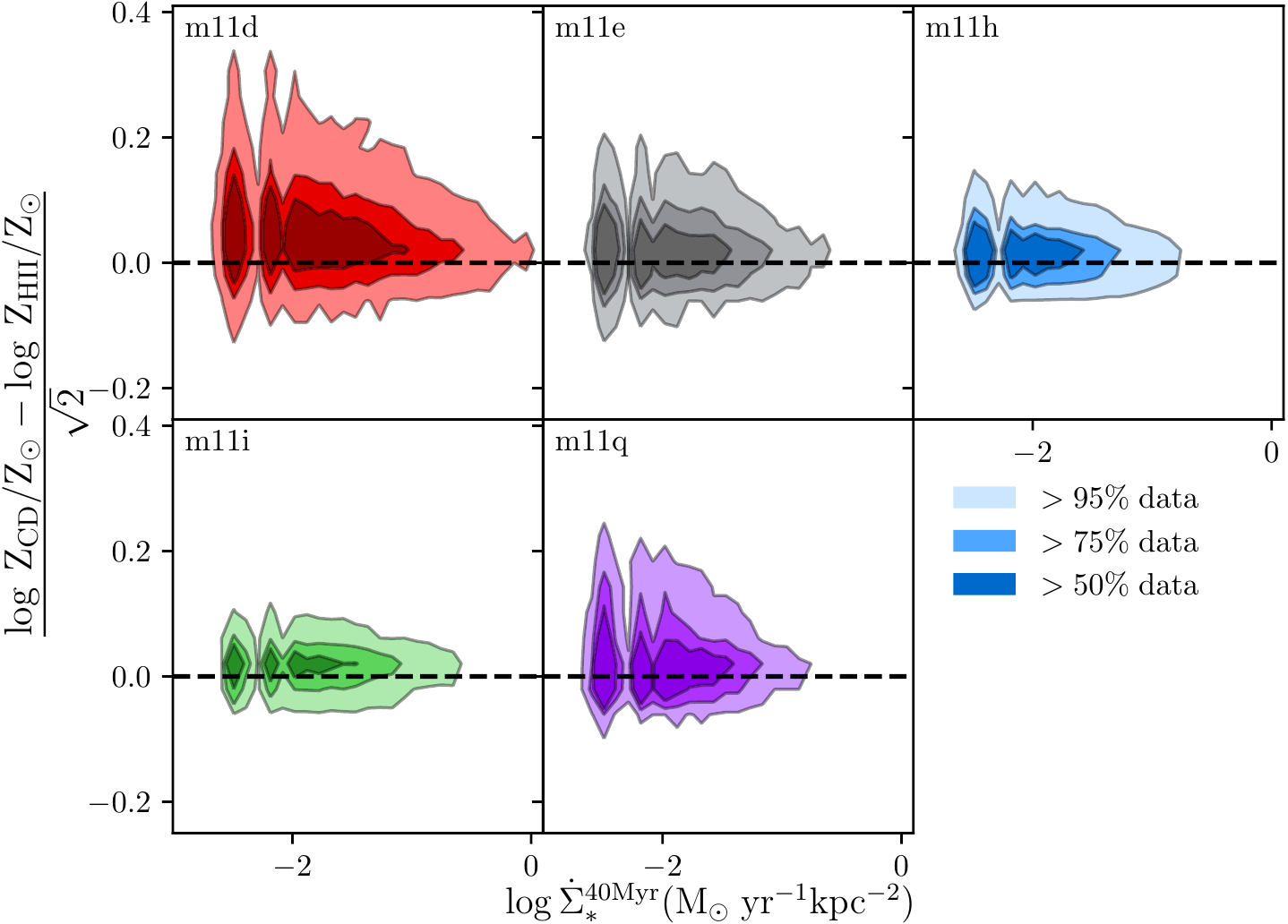}
    \caption{Comparing the distribution of star formation rate surface density averaged over the last 40 Myrs to the DED for all five galaxies, in the style of Figure~\ref{fig:contours_masscut_CDMassvsD}. Filled contours denote $50\%, 75\%$, and $95\%$ of data inclusion, including diffuse components. Colors as Figure~\ref{fig:SFR_Mstar_overtime}. All distributions appear to be unbiased, with the exception of m11d. No general dependence on the differential enrichment as a function of SFR is seen in any of the simulations.}
    \label{fig:contours_SFRfortyvsD}
\end{figure*}

We show the distribution of the DED in each of the galaxies, as a function of gas surface density and star formation rate surface density, in Figures~\ref{fig:contours_masscut_CDMassvsD} and \ref{fig:contours_SFRfortyvsD}. We conclude that the cold and dense gas appears to almost always be slightly more enriched than the ionized gas on average. We find no significant correlation with the Differential Enrichment Distance and other physical factors such as velocity dispersion, gas surface density, and star formation rate. 

We note that for the nebular regions there does not appear to be a significant relationship for the cold and dense gas surface density or 40Myr SFR. We see a slight striation pattern in the latter (purely an effect of resolution, i.e., counting individual young star particles), but neither seem to show any significant relationships aside from the fact that cold and dense gas is slightly more enriched, despite the SFR and gas surface density.

\section{Relating Metallicity to Star Formation} \label{Z_and_SFR}

\begin{figure*}
    \includegraphics[width=\textwidth]{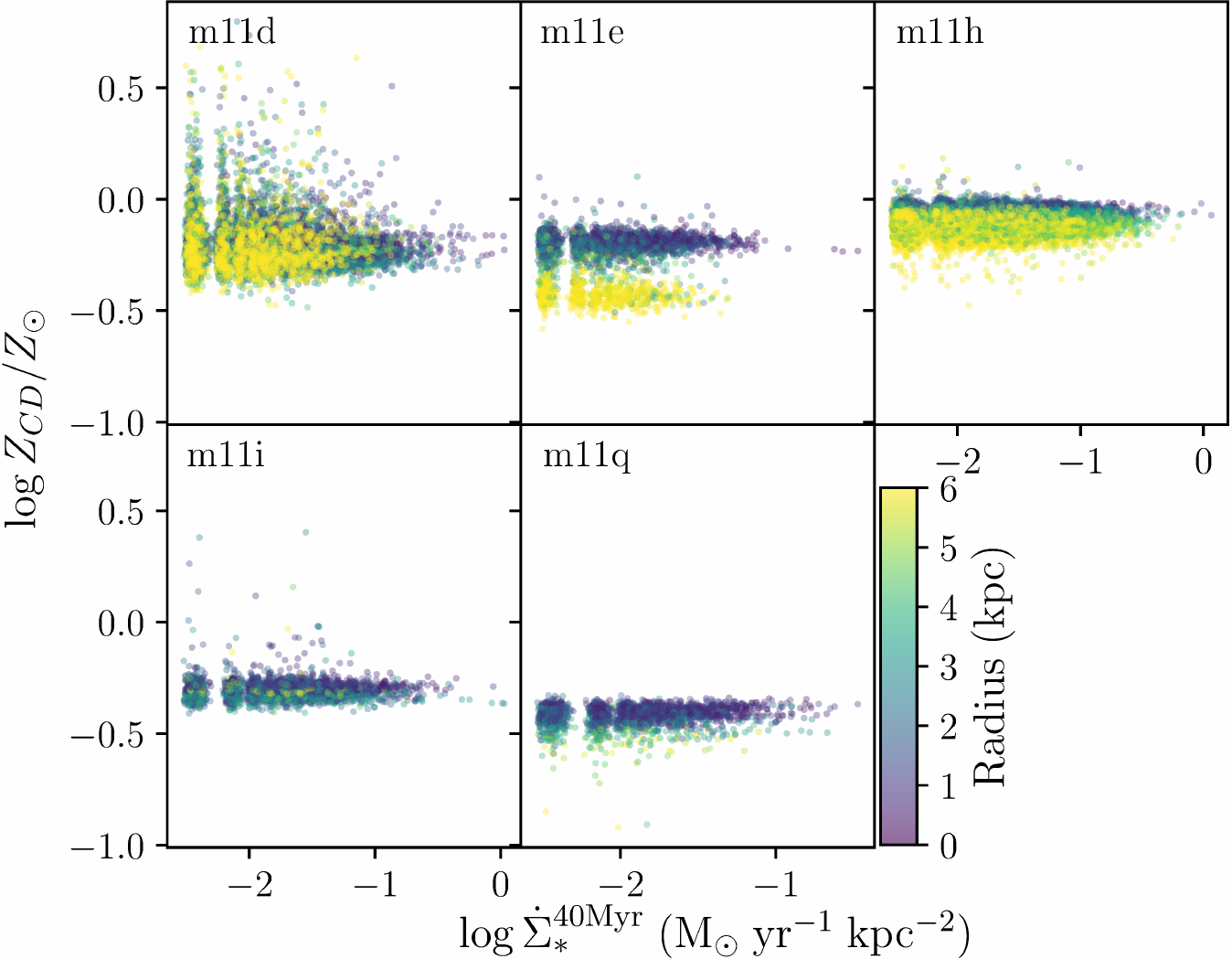}
    \caption{Distribution of the log star formation rate averaged over the last 40 Myrs as surface density (per square parsec) compared to the total solar metallicity in the cold and dense gas for all five galaxies, from all snapshots. The color of pixels is determined by the radius in kpc. All distributions appear to have little-to-no correlation between the quantities.}
    \label{fig:scatter_SFRfortysdvsZ}
\end{figure*}

During this study, an effort was made to correlate the local star formation rate, metallicity, and effective radius in these five FIRE-2 dwarf galaxies. This can be found in Figure~\ref{fig:scatter_SFRfortysdvsZ}, where we represent the star formation rate's surface density (averaged over the last 40 Myr, in units of $M_{\odot} yr^{-1} pc^{-2}$), plotted against the log of the total solar metallicity in the cold and dense gas. There appears to be no relationship in any of the five galaxies between these two quantities, with no further dependence on the effective radius as well. We note that while galaxy m11e appears to have a smaller, mirrored distribution at lower metallicity and higher radius, that this is likely from the snapshots of the merging companion, which intuitively explain the high radius and low metallicity values. 

\bsp	
\label{lastpage}
\end{document}